%% file: article1201.tex
\documentclass[12pt]{article}

\usepackage{amsmath,amsthm,amsfonts,amscd,amssymb,epsfig}
\usepackage[all]{xy}

\include{notations}

\begin{document}

\newenvironment{rem}
{\addtocounter{rem}{1}\setlength{\topsep}{1em}\par\trivlist\item{\bf Remark
\arabic{section}.\arabic{rem}.} }{\endtrivlist}

\newenvironment{ass}
{\addtocounter{ass}{1}\setlength{\topsep}{1em}\par\trivlist\item{\bf Assumption
\arabic{section}.\arabic{ass}.} }{\endtrivlist}

\newenvironment{ex}
{\addtocounter{ex}{1}\setlength{\topsep}{1em}\par\trivlist\item{\bf Example
\arabic{section}.\arabic{ex}.} }{\endtrivlist}

\newenvironment{demo}[1][Proof]
{\setlength{\topsep}{1em}\par\trivlist\item{\em #1.} }
{\openbox\endtrivlist}

\newenvironment{cond}{\begin{list}{{\it (\roman{ploum})}}{\usecounter{ploum}}}
{\end{list}}

\title{Moments and Cumulants of 
Polynomial random variables on unitary groups,
the Itzykson Zuber integral and free probability}
\author{Beno\^\i{}t Collins}
\date{D.M.A, Ecole Normale Supérieure\\
45, rue d'Ulm, 75005 Paris, France \\
{\tt benoit.collins@ens.fr}\\ fax: +33 1 44 32 20 80}

\maketitle

\begin{abstract}
We consider integrals on unitary groups $\mathbb{U}_d$ of the form
\begin{equation*}
\int_{\mathbb{U}_{d}}U_{i_{1}j_{1}}\cdots U_{i_{q}j_{q}}U^{*}_{j'_{1}i'_{1}}\cdots
U^{*}_{j'_{q'}i'_{q'}}dU
\end{equation*}
We give an explicit formula in terms of characters of symmetric groups and
Schur functions, which allows us to rederive an asymptotic expansion as
$d\rightarrow\infty$. Using this we rederive and strenghthen a result of 
asymptotic freeness due to Voiculescu.

We then study large $d$ asymptotics of matrix model integrals 
and of the logarithm of Itzykson-Zuber integrals and show that they converge
towards a limit when considered as power series.
In particular we give an explicit formula for 
\begin{equation*}
\lim_{d\rightarrow\infty}\frac{\partial^n}{\partial z^n}d^{-2}\log\int_{\mathbb{U}_d}
e^{zd\Tr (XUYU^*)}dU|_{z=0}
\end{equation*}
assuming that the normalized traces $d^{-1}\Tr(X^k)$ and $d^{-1}\Tr (Y^k)$
converge in the large $d$ limit.
We consider as well a different scaling and relate its asymptotics
to Voiculescu's R-transform.
\end{abstract}

\section{Introduction}\label{intro}

The starting point of this paper is a recent result by Guionnet and
Zeitouni (see \cite{MR1883414}).
For all integers $d$, let $X_d$ and $Y_d$ be hermitian matrices in $\M{d}$
such that for all $k\geq 0$, $\lim_d d^{-1}\Tr (X_d^k)$ (resp.
$\lim_d d^{-1}\Tr (Y_d^k)$) exists in $\mathbb{R}$ and equals $x_k$ 
(resp. $y_k$). Assume furthermore that there exists a majorant $A$ independent
on $d$ for all $||X_d||$ and $||Y_d||$. Consider the following integral on
the unitary group $U_d$ with its normalized Haar measure $dU$:
\begin{equation*}
F_d=d^{-2}\log\int_{\mathbb{U}_d}e^{d\Tr (X_dUY_dU^*)}dU
\end{equation*}
It is proved in \cite{MR1883414}, that 
$\lim_d F_d$ exists and depends only on the sequences $(x_k)$ and $(y_k)$.

This result, whose proof implies subtle computations and results about large deviation
bounds for Brownian motion, raises a number of questions. For example,
it is of interest 
to understand what would happen if the matrices fail to be hermitian or
to have a norm bounded independently on $d$.

Our approach to this problem is based on an explicit algebraic computation of the
integral of polynomial functions on unitary groups. This approach yields
a weaker convergence concept than that of \cite{MR1883414} and the results of this paper
don't allow to recover the results of Guionnet and Zeitouni. On the bright side,
we show that a very large class of integrals similar to $F_d$ converge 
in our sense and we obtain an explicit expression for their limit.

With \cite{MR57:11421}, Weingarten was among the first try to understand
the integral of polynomial functions on unitary groups for large $d$. 
We supply a new proof of Weingarten's result and generalize it by 
giving an explicit way of computing an integral of a
polynomial random variable on the unitary group $\mathbb{U}_d$ with characters and
Schur functions. This is the content of Theorem \ref{thWe} of Section \ref{s-2}.
In theorems \ref{combWe} and \ref{DLexplicite},
we give a way of computing these integrals and cumulants associated to
them by summing over sequences of permutations satisfying specific properties.

Section \ref{s-3} is an application of theorem \ref{thWe} to 
a new proof (theorem \ref{libre}) and an improvement (theorem \ref{proba}) of an 
asymptotic freeness result of Voiculescu and Feng Xu. 
For a set $W$ and for each $d$, let $(w_{i,d})_{i\in W}$ be a family of 
matrices in $\M{d}$ such that for any non-commutative polynomial
$P$ in $W$, $\lim_d (d^{-1}\Tr (P(w_{i,d})))$ admits a finite limit
in $\mathbb{C}$.

Let $U_1,U_2,\ldots $ be a sequence of Haar distributed unitary independent matrices.
Take the convention that $\tr=d^{-1}\Tr$.
\begin{Thm*}
\begin{itemize}
\item (i)
For any non-commutative polynomial $Q$ in the variables 
$W, U_1, U_1^*,U_2,U_2^*\ldots$, $d^{-1}\Tr (Q(w_{i,d},U_1^*,U_2,U_2^*\ldots ))$
admits a finite limit in $\mathbb{C}$.
\item (ii)
The family of sets of variables $W,\{U_1,U_1^*\},\ldots $ is asymptotically free.
\item (iii)
Furthermore, we have
\begin{equation*}
P(|\tr (Q(w_{i,d},U_1,U_1^*\ldots ))
-\lim_d E (\tr (Q(w_{i,d},U_1,U_1^*\ldots )))|\geq\varepsilon) =O(d^{-2})
\end{equation*}
\end{itemize}
\end{Thm*}

The two first points are stated in theorem \ref{libre} and are due to Feng Xu
\cite{MR99f:81185}. We supply a new proof of these results. The third point is
the object of theorem \ref{proba}.
The methods gathered in this section are used for section \ref{s-4}, in which
we first show the convergence theorem  \ref{universal}:


\begin{Thm*}
Let $(P_{l,j})_{1\leq l \leq k , 1\leq j \leq k}$ 
and
$(Q_{l,j})_{1\leq l \leq k , 1\leq j \leq k}$ be two families of noncommutative polynomials
in $U_1,U_1^*,\ldots ,U_k,U_k^*$ and $W$.
Let $A_d$ be the variable $\sum_{l=1}^k\prod_{j=1}^k\tr P_{l,j}(U,U^*,w_{i,d})$ and
$B_d$ the variable $\sum_{l=1}^k\prod_{l=1}^k\tr Q_{i,j}(U,U^*,w_{i,d})$.
\begin{itemize}
\item
The function 
\begin{equation*}
z\rightarrow d^{-2}\log E (\exp (zd^2 A_d)) = \sum_{q\geq 1} a_q^d z^d
\end{equation*}
is such that for all $q$, $\lim_d a_q^d$ exists and depends only on $P_{l,j}$
and the limit distribution of $W$.
\item
The function 
\begin{equation*}
z\rightarrow \frac{E \exp (zB_d+zd^2 A_d)}{E \exp zd^2 A_d} = 1+\sum_{q\geq 1} b_q^d z^d 
\end{equation*}
is such that for all $q$, $\lim_d b_q^d$ exists and depends only on 
$P_{l,j},Q_{l,j}$ and the limit distribution of $W$.
\end{itemize}
\end{Thm*}

The explicit link with the work of Guionnet and Zeitouni is the following:
let $F$ be the function defined by
\begin{equation*}
F_{d,X,Y}=d^{-2}\log E (\exp zd^2 \tr (XUYU^*))
\end{equation*}
where $X$ and $Y$ are arbitrary $d\times d$ matrices.
Then $\frac{\partial^q}{\partial z^q}F_{d,X,Y}(0)$ is a polynomial function of the variables
$\tr X^i$ and $\tr Y^i$, whose coefficients are rational fractions of $d$.
We prove that these coefficients converge as $d\rightarrow\infty$ to
some limit, for which we give an explicit formula 
(theorem \ref{Fborne}) and a diagrammatic interpretation 
(theorem \ref{diagr}).

We finish this paper by establishing in theorem \ref{thmfree}
the following link between 
free probability and the Itzykson-Zuber integral

\begin{Thm*}
Assume that $X$ is a one dimensional projector and that $Y$ admits
a limit distribution.
Then for all $q$, the number
$d\cdot \partial^q/\partial z^q F_{d,X,Y}(0)$ converges towards
$(q-1)!k_q(Y)$, i.e. the coefficient of the primitive
of Voiculescu's $R$ -transform $R_Y$ 
of the limit distribution of $Y$. 
\end{Thm*}

This article is a part of the author's Ph.D. work.
The author would like to thank his advisor Philippe Biane for numerous discussions and
for introducing him to the subjects to which this paper is related. He also
thanks Paul Zinn-Justin for useful conversations.

\section{Integrals  of polynomial functions   
on $\mathbb{U}_d$}\label{s-2}

\subsection{An exact formula for Weingarten's function}\label{ss-2.1}

Let $\mathbb{U}_d$ be the group of unitary $d\times d$ matrices and $dU$ be its
normalized Haar measure.
 Let $q,q'$ be a positive integers and
$\mathbf{i}=(i_1,\ldots ,i_q)$, $\mathbf{i'}=(i'_1,\ldots ,i'_{q'})$,
$\mathbf{j}=(j_1,\ldots ,j_q)$, $\mathbf{j'}=(j'_1,\ldots ,j'_{q'})$
be two $q$ -uples and two $q'$-uples of indices in $[1,d]$, we define
\begin{equation*}
I_{d,\mathbf{i,i',j,j'}}=
\int_{\mathbb{U}_{d}}U_{i_{1}j_{1}}
\cdots U_{i_{q}j_{q}}U^{*}_{j'_{1}i'_{1}}\cdots
U^{*}_{j'_{q'}i'_{q'}}dU
\end{equation*} 
We shall give an explicit formula for this integral in terms of Schur functions
and characters of symmetric groups.
Using the invariance of Haar measure under multiplication by scalar unitary
matrices one checks that this integral is zero if $q\neq q'$ so
we will only consider the case $q=q'$.

In order to state our result we need to introduce some notations. 
Let $(E_{ij})_{i,j\in[1,d]}$ be the canonical basis of $\M{d}$, we define
\begin{equation*}
E_{d,\mathbf{i,i',j,j'}}:=E_{i_1i'_1}\otimes\cdots \otimes E_{i_qi'_q}\otimes
E_{j_1j'_1}\otimes\cdots \otimes E_{j_qj'_q}\in\M{d}^{\otimes q}\otimes
\M{d}^{\otimes q}
\end{equation*}
and the linear 
form $I_{d,q}$ on $\M{d}^{\otimes q}\otimes
\M{d}^{\otimes q}$
 such that $I_{d,q} ( E_{d,\mathbf{i,i',j,j'}})= I_{d,\mathbf{i,i',j,j'}}$.
We denote $\Sy{q}$ the symmetric group on $\{1,\ldots,q\}$.
For $\sigma,\tau\in\Sy{q}$
we define  $\delta_{(\sigma, \tau)}$ as the  linear form on 
$\M{d}^{\otimes q}\otimes
\M{d}^{\otimes q}$ such that
\begin{equation*}
\delta_{(\sigma, \tau)}(E_{d,\mathbf{i,i',j,j'}}):=
\delta_{i_{1}i'_{\sigma(1)}}\cdots\delta_{i_{q}i'_{\sigma(q)}}
\delta_{j_{1}j'_{\tau(1)}}\cdots\delta_{j_{q}j'_{\tau(q)}}
\end{equation*}

We shall follow the standard notations concerning partitions, Schur functions,
and characters of symmetric groups, namely if $\lambda\vdash q$, i.e. $\lambda$
is a partition of $q$, we denote by 
$\chi^{\lambda}$ the corresponding character of $\Sy{q}$ and by
$s_{\lambda ,d}(x_1,\ldots, x_d)$ the Schur function, see
 \cite{MR99f:05119} or \cite{MR96h:05207}.
Whenever convenient we shall let
$s_{\lambda ,d}(x)=s_{\lambda ,d}(x,\ldots ,x)$. 
If $\mu\vdash q$ we denote $C_{\mu}$ the corresponding conjugacy class of
$\Sy{q}$, and $Z_{\mu}$ the number of elements of $C_{\mu}$.
 Finally, if $\sigma\in\Sy{q}$ we
denote by $|\sigma|$ the minimal number $k$ such that $\sigma$ can be written as a
product of $k$ transpositions. Recall that $|\sigma|=q-c(\sigma)$, where
$c(\sigma)$ is the number of cycles of $\sigma$.

We can now state the main result of this section.
\begin{Thm}\label{thWe}
Let $q,d$ be integers satisfying $d\geq q$; one has
\begin{equation}\label{eqWe}
I_{d,q}=\sum_{\sigma, \tau \in \Sy{q}} \delta_{(\sigma,\tau)}
\frac{1}{q!^2}\sum_{\lambda\vdash q}
\frac{\chi^{\lambda}(1)^2\chi^{\lambda}(\sigma\tau^{-1})}{s_{\lambda ,d}(1)}
\end{equation}
\end{Thm}

We shall denote $\We$ the function of $d,q,\sigma\tau^{-1}$ which occurs in the above 
result. For integers $d,q$ and $\sigma\in\Sy{q}$ one has
\begin{equation}\label{defWe}
\We(d,q,\sigma)=\frac{1}{q!^2}\sum_{\lambda\vdash q}
\frac{\chi^{\lambda}(1)^2\chi^{\lambda}(\sigma)}{s_{\lambda ,d}(1)}
\end{equation}
Observe that $s_{\lambda ,d}(1)$, which is the dimension of the irreducible
representation of $\mathbb{U}_d$ associated with $\lambda$, is a polynomial function of
$d$, of degree $q$, therefore, 
$\We$ is a rational function of $d$, of degree at most $-q$.
Since the dependence of $\We$ in $q$ is implicitly given by $\sigma$, we shall
drop it from notations and use $\We(d,\sigma)$ when it causes no confusion.
\begin{demo}

We consider the left action $\pi$ of $\mathbb{U}_d$ on $\M{d}$ by conjugation, i.e. 
\begin{equation*}
\pi(U)M:=UMU^*
\end{equation*}
and $\pi^{\otimes q}$ the corresponding $q$ -fold tensor product
action of $\mathbb{U}_d$ on $\M{d}^{\otimes q}$. 
Then we consider the action $\pi^{\otimes q}\otimes\pi^{\otimes q}$ of 
$U_d\times U_d$ on $W=\M{d}^{\otimes q}\otimes \M{d}^{\otimes q}$.

The linear form $\delta_{(\sigma, \tau)}$ is invariant under the action of 
$\mathbb{U}_d\times \mathbb{U}_d$. By  Schur-Weyl duality, the collection
$\{ \delta_{(\sigma, \tau)} \}$ is a generating family of the vector space of
$\mathbb{U}_d\times \mathbb{U}_d$-linear maps, furthermore 
it is a basis for $d\geq q$, so that one can write
\begin{equation*}
I_q=\sum_{\sigma, \tau \in S_{q}} 
\delta_{(\sigma, \tau)}
\cdot\Gamma (d,\tau, \sigma )
\end{equation*}
and the coefficients $\Gamma (d,\tau, \sigma )$ are uniquely defined
for $d \geq q$.
Let $\Phi_{\sigma}$ be the linear endomorphism of 
$\M{d}^{\otimes q}\otimes \M{d}^{\otimes q}$ defined by
\begin{equation} 
\begin{split}
\Phi_{\sigma}(E_{i_1i'_1}\otimes\cdots \otimes E_{i_qi'_q}\otimes
E_{j_1j'_1}\otimes\cdots \otimes E_{j_qj'_q}):= \\
E_{i_1i'_{\sigma (1)}}\otimes\cdots \otimes E_{i_qi'_{\sigma (q)}}\otimes
E_{j_1j'_{\sigma (1)}}\otimes\cdots \otimes E_{j_q j'_{\sigma (q)}} \\
\end{split}
\end{equation}
For all $\sigma\in\mathcal{S}_q$ one has
$I_q\circ\Phi_{\sigma}=I_q$. On the other hand one has
$\delta_{(\sigma, \sigma ')}\circ
\Phi_{\tau}=\delta_{(\sigma \tau, \sigma '\tau)}$.
This proves that $\Gamma (d,\sigma, \sigma ')= 
\Gamma (d,\sigma\tau, \sigma '\tau)$. 
In the same way, defining
\begin{equation} 
\begin{split}
\Phi_{\sigma}'(E_{i_1i'_1}\otimes\cdots \otimes E_{i_qi'_q}\otimes
E_{j_1j'_1}\otimes\cdots \otimes E_{j_qj'_q}):= \\
E_{i_{\sigma (1)}i'_1}\otimes\cdots \otimes E_{i_{\sigma (q)}i'_q}\otimes
E_{j_{\sigma (1)}j'_1}\otimes\cdots \otimes E_{j_{\sigma (q)} j'_q} \\
\end{split}
\end{equation}
gives that $\Gamma (d,\sigma, \sigma ')= 
\Gamma (d,\tau\sigma, \tau\sigma ')$. This proves that $\Gamma$ 
only depends on the conjugacy class of $\tau\sigma^{-1}$.
We call $\We(d,\tau\sigma^{-1})$ the function such that 
$\Gamma (d,\sigma,\tau)=\We(d,\tau\sigma^{-1})$.

Note that choosing 
$\mathbf{i}=(1,\ldots ,q)$, $\mathbf{i'}=(1,\ldots ,q)$, 
$\mathbf{j}=(1,\ldots ,q)$ and $\mathbf{j'}=(\sigma(1),\ldots ,\sigma(q))$
one finds that
\begin{equation}
\We (d,\sigma)=\int_{\mathbb{U}_d}U_{11}\ldots U_{qq}\overline{U_{1\sigma(1)}}
\ldots \overline{U_{q\sigma(q)}}dU
\end{equation}
Consider now 
\begin{equation*}
 A=\bigotimes_{i=1}^{q}E_{ii}\in\M{d}^{\otimes q}\quad\text{and}\quad
 B=\bigotimes_{i=1}^{q}E_{i\sigma (i)}\in\M{d}^{\otimes q}
\end{equation*}
Denoting  $\rho (U)$ the 
matrix $U\otimes \ldots \otimes U\in\M{d}^{\otimes q}$ one has
\begin{equation*}
\We(d,\sigma)=\int_{\mathbb{U}_d}(\Tr (A \rho (U) B \rho (U)^{*}))dU
\end{equation*}
For $\tau\in\Sy{q}$ let
$A_{\tau}=\bigotimes_{i=1}^{q}E_{\tau (i) \tau (i)}$ and
$B_{\tau}=\bigotimes_{i=1}^{q}E_{\tau (i) \tau (\sigma (i))}$.
The unitary permutation matrix $U_{\tau}=\sum_{i,j}E_{ij}\delta_{i\tau (j)}\in\M{d}$ 
satisfies
$A_{\tau}=\rho(U_{\tau})A\rho(U_{\tau}^*)$ and  
$B_{\tau}=\rho(U_{\tau})B\rho(U_{\tau}^*)$ so that, since
$dU$ is a Haar measure, we have, for $\tau,\tau '\in\Sy{q}$,
\begin{equation*}
\int_{\mathbb{U}_d}(\Tr (A_{\tau} \rho (U) B_{\tau '} \rho (U)^{*})dU=
\int_{\mathbb{U}_d}(\Tr (A \rho (U) B \rho (U)^{*})dU
\end{equation*}
Let
$\tilde{A}=\sum_{\tau\in\Sy{q}}
\bigotimes_{i=1}^{q}E_{\tau (i) \tau (i)}$ and
$\tilde{B}=\sum_{\tau\in\Sy{q}}
\bigotimes_{i=1}^{q}E_{\tau (i) \tau (\sigma (i))}$,
we have thus 
\begin{equation}\label{rhs}
q!^2\int_{\mathbb{U}_d}(\Tr (A \rho (U) B \rho (U)^{*}))dU=
\int_{\mathbb{U}_d}(\Tr (\tilde{A} \rho (U) \tilde{B} \rho (U)^{*})dU
\end{equation}
We  compute  the r.h.s. of \ref{rhs}.
Let $\Psi$ be the action of $\Sy{q}$ on
$\M{d}^{\otimes q}$ defined by
$$\Psi(\sigma) E_{i_1j_1}\otimes\ldots\otimes E_{i_qj_q}=
E_{i_{\sigma (1)}j_{\sigma (1)}}\otimes\ldots\otimes 
E_{i_{\sigma (q)} j_{\sigma (q)}}.$$
Consider the algebra of elements of $\M{d}^{\otimes q}$ fixed under the action
$\pi^{\otimes q}\otimes\Psi$ of $\mathbb{U}_d\times \Sy{q}$.
 It is known from representation theory
that this algebra is abelian and that its minimal projections are indexed by
$\lambda\vdash q$. We denote by
$\{ I_{\lambda} \}_{\lambda\vdash q}$ 
these minimal projections. The dimension of the range of $I_{\lambda}$ is
$\chi^{\lambda}(e)s_{\lambda ,d}(1)$. 
One has  
$\Tr (\tilde{B}I_{\lambda})=\chi^{\lambda}(e)\chi^{\lambda}(\sigma)$, therefore 
 $\int \rho (U)\tilde{B}I_{\lambda}\rho (U^*)dU$, which is
is a multiple of $I_{\lambda}$, is fully determined
by its trace. thus,
\begin{equation}
\int_{\mathbb{U}_d}\rho (U)\tilde
{B}I_{\lambda}\rho (U^*)dU=\frac{\chi^{\lambda}(e)\chi^{\lambda}(\sigma)}
{s_{\lambda ,d}(1)\chi^{\lambda}(e)}I_\lambda
\end{equation}
Multiplying by $\tilde A$ and taking the trace yields
\begin{equation}
\int_{\mathbb{U}_d}(\Tr (\tilde{A} \rho (U){B}I_{\lambda} \rho (U)^{*}))=
\frac{\chi^{\lambda}(e)^2\chi^{\lambda}(\sigma )}{s_{\lambda ,d}(1)}
\end{equation}

Summing over $\lambda\vdash q$ gives the result.
\end{demo}

\subsection{Asymptotic expansion for $\We$}
Remember from the remark following the statement of
Theorem \ref{thWe} that $\We$ is a rational function of $d$,
of degree at most $-q$,
therefore it has a Laurent expansion 
\begin{equation}\label{eqcombWe}\We(d,\sigma)=\sum_{l=0}^{\infty}
A[\sigma,l]d^{-q-l}
\end{equation}
Since the roots of the polynomials $s_{\lambda ,d}(1)$ belong to the set
$\{0,1,\ldots,q-1\}$ we know that the radius of convergence of the series
is at least $1/(q-1)$, hence  \ref{eqcombWe}  converges for all $d\geq q$.
We now determine the coefficients of this expansion.
\begin{Thm}\label{combWe}
Let $\sigma\in\Sy{q}$.
\item (i)
One has $A[\sigma,0]=1$ if $\sigma=e$ and $A[\sigma,0]=0$ for $\sigma\not=e$.
\item (ii)
For $l\neq 0$, let $A[\sigma, k, l]$ be the number
of $k$-tuples of permutations $(\sigma_1, \ldots, \sigma_k)$, each different 
from the identity, satisfying both $\sum_{i=1}^k|\sigma_i|=l$ 
and $\sigma\sigma_1\ldots\sigma_k=e$, then  one has 
 $$A[\sigma,l]=\sum_{k=1}^l (-1)^k A[\sigma,k,l].$$
\end{Thm}

Before we prove this result, we deduce a useful corollary.

\begin{Cor}\label{asWe}
For any $\sigma\in\Sy{q}$, the rational 
function $d^{q+|\sigma |}\We(d,\sigma)$ is  even, and its degree is less or equal than
 $0$.
\end{Cor}

\begin{demo}
If $\sum_{i=1}^k |\sigma_i |=l < |\sigma |$, then by the triangle inequality
$|\sigma\sigma_1\ldots\sigma_k|>0$, in other words the product
$\sigma_1\ldots\sigma_k$ can not be $e$.
So $A[\sigma ,l]=0$.
This proves the assertion regarding the degree.
Let $\varepsilon(\sigma)\in\{-1,1\}$ denote the sign of a permutation $\sigma$.
 If $\sigma\sigma_1\ldots\sigma_k=e$ then 
$\varepsilon(\sigma\sigma_1\ldots\sigma_k)=1=\varepsilon(\sigma)
\varepsilon(\sigma_1)\cdots\varepsilon(\sigma_k)$
Since for every  $\tau$ one has 
$\varepsilon(\tau)=(-1)^{|\tau |}$ we see that $A[\sigma,l]=0$
if $l$ and $|\sigma |$ do not have the same parity.
\end{demo}

In order to prove Theorem \ref{combWe} we shall need some preliminary results. 
First we introduce some
notation. Let $k$ be an integer and $\mu_0,\mu_1,\ldots,\mu_k$
 be partitions of $q$. Let $\sigma\in\Sy{q}$ be such that $\sigma\in C_{\mu}$.
 We
call $A[\mu ; \mu_1,\ldots ,\mu_k]$ the number of $k$-tuples
$(\sigma_1,\ldots,\sigma_k)\in \Sy{q}$ such that  
$\sigma\cdot\sigma_1\cdots\sigma_k=e$ and $\sigma_i\in C_{\mu_i}$
for all $i\in [1,k]$. 

\begin{Lem}\label{lemcentre}
Let $\mu,\mu_1,\ldots,\mu_k$ be partitions of $q$ and $\sigma\in C_{\mu}$
then
$$\frac{1}{q!}\sum_{\lambda}\chi^{\lambda}(\sigma)\chi^{\lambda}_{\mu_1}\ldots
\chi^{\lambda}_{\mu_k}(\chi^{\lambda}(e))^{1-k}\cdot
Z_{\sigma_1}\ldots Z_{\sigma_k}=A[\mu;\mu_1,\ldots,\mu_k].$$
\end{Lem}
\begin{demo}A standard computation in the group algebra of $\Sy{q}$, using
Fourier analysis.
\end{demo}

Next we recall another well known formula, see e.g.
\cite{MR99f:05119} or \cite{MR96h:05207}.
~\

\begin{Lem}\label{sl1}
For any $\lambda\vdash q$ one has
 $s_{\lambda ,d} (1)=\frac{1}{q!}\sum_{\tau\vdash q}
  d^{q-|\tau |}\chi^{\lambda }(\tau )Z_{\tau}$.
\end{Lem}

~\
We now  proceed to the proof of Theorem \ref{combWe}.

\begin{demo}
From Lemma \ref{sl1} one has
\begin{equation*}
s_{\lambda ,d}(1)=\frac{d^q}{q!}(1+\sum_{\tau\vdash q, \tau\not=1^q}
d^{-|\tau|}\chi^{\lambda}(\tau)Z_{\tau})
\end{equation*}
Expand $1/s_{\lambda ,d}(1)$ in the expression
\begin{equation*}
\We(d,\sigma )=\frac{1}{d^qq!}
\sum_{\lambda\vdash q}\frac{\chi^{\lambda}(\sigma)
\chi^{\lambda}(e)}{\sum_{\tau \vdash q}d^{-|\tau|}\chi^{\lambda}(\tau)
Z_{\tau}/\chi^{\lambda}(e)}
\end{equation*}
to find that the coefficient of $d^{-q-l}$ is
\begin{equation}
\begin{split}
\frac{1}{q!}\sum_{\lambda\vdash q}\chi^{\lambda}(\sigma)
\chi^{\lambda}(e)\sum_{k=1}^l (-1)^k \sum_{\mu_1,\ldots ,\mu_k\vdash q,
 \sum_{i=1}^k |\mu_i|=l}
\prod_{i=1}^k\frac{\chi^{\lambda}_{\mu_i}}{\chi^{\lambda}(e)}Z_{\sigma_i}\\
=\frac{1}{q!}\sum_{k=1}^{l} (-1)^k
\sum_{\mu_1,\ldots ,\mu_k\vdash q,
 \sum_{i=1}^k |\mu_i|=l}
\sum_{\lambda\vdash q}\frac{\chi^{\lambda}(\sigma )\chi^{\lambda}(\sigma_1 )
\ldots \chi^{\lambda}(\sigma_k )}{(\chi^{\lambda}(e))^{k-1}}
\cdot Z_{\sigma_1}\ldots Z_{\sigma_k}\\
\end{split}
\end{equation}

Using Lemma \ref{lemcentre}, this expression is seen to be
\begin{equation*}
\sum_{k=1}^{l} (-1)^k\sum_{\mu_1,\ldots ,\mu_k\vdash q,
 \sum_{i=1}^k |\mu_i|=l}
A[\sigma ; \mu_1,\ldots ,\mu_k]
\end{equation*}
From the definition of $A[\sigma,k,l]$ we have
\begin{equation*}
\sum_{\mu_1,\ldots ,\mu_k\vdash q,
 \sum_{i=1}^k |\mu_i|=l}
A[\sigma ; \mu_1,\ldots ,\mu_k]=A[\sigma,k, l]
\end{equation*}
which concludes the proof.
\end{demo}

\subsection{Cumulants of random variables}\label{crv}

For reference and proofs about the beginning of this section, we refer to
\cite{MR98m:81037} or \cite{MR30:4688}.
We consider the lattice $\mathcal{P}$ of partitions of $[1,q]=\{1,\ldots ,q\}$, and
say that $\Pi'\leq\Pi$ if, for any block $V$ of $\Pi'$ there exists a block
of $\Pi$ that contains $V$. For any
partitions $\Pi$ and $\Pi'$ define $\Pi\vee\Pi'$ (resp. $\Pi\wedge\Pi'$) to be
the smallest majorant (resp. the greatest minorant), and
the greatest element $1_q=\{\{1,\ldots ,q\}\}$ 
(resp. $0_q=\{\{ 1\},\ldots ,\{ q\}\}$) the greatest (resp. smallest) element.

The permutation group $\Sy{q}$ admits a natural left action
on the partitions of $[1,q]$. Call
$\mathcal{O}_{\Pi}$ the orbit to which $\Pi$ belongs. These orbits
are in natural one to one correspondence with the partitions of the
integer $q$.
For $\sigma\in\Sy{q}$, let $\Pi_{\sigma}$ be the partition of $[1,q]$ whose
blocks are the orbits of $\sigma$.
It is obvious but useful to remark that 
$C_{\sigma}=\mathcal{O}_{\Pi_{\sigma}}$.

For each $\Pi_1,\Pi_2\in\mathcal{P}$ such that $\Pi_1\leq\Pi_2$, 
there exists a partition $\Pi$ such that
the interval $[\Pi_1 ,\Pi_2]$ is isomorphic as a lattice to the lattice
$[0_q,\Pi]$. If $\mathcal{O}_{\Pi}=(p_1,\ldots,p_k)$ then define
\begin{equation*}
\Moeb (\Pi_1 ,\Pi_2)=\prod_{i} ((-1)^{i-1}(i-1)!)^{p_i}
\end{equation*}
For a set $A_1,\ldots,A_q$ of random variables and 
$\Pi=\{V_1,\ldots ,V_k \}$ a partition
of $[1,q]$, let
\begin{equation*}
E_{\Pi}(A_1,\ldots ,A_q)=\prod_{i=1}^kE(\prod_{j\in V_i} A_j)
\end{equation*}
With the same notations we define the {\it classical cumulant \it}
by
\begin{equation*}
C_{\Pi}(A_1,\ldots ,A_q)=\sum_{\Pi'\leq \Pi}E_{\Pi '}(A_1,\ldots ,A_q)
\Moeb(\Pi',\Pi)
\end{equation*}
and for $\Pi=1_q$ we define $C_q(A_1,\ldots ,A_q)=
C_{\Pi}(A_1,\ldots ,A_q)$. 

We also introduce the notion of {\it relative cumulant \it}
\begin{equation}\label{cumrel}
C_{\Pi_1,\Pi_2}(A_1,\ldots ,A_q)=\sum_{\Pi_1\leq \Pi\leq\Pi_2}E_{\Pi }(A_1,\ldots ,A_q)
\Moeb (\Pi,\Pi_2)
\end{equation}
This cumulant is the cumulant of the function $E_{\Pi}$ associated to the lattice
$[\Pi_1,1_q]$. 
It is known that $\sum_{q\geq 1}C_q(A,\ldots ,A)z^q/q!$
 is the power series expansion in a neighbourhood
of $0$ of the analytic function $z\rightarrow \log E(\exp (zA))$. 
Let $\Pi_1\leq\Pi_2$ be partitions of $[1,q]$  
and $A_1,\ldots A_q$ be random variables. We have
\begin{equation}\label{rota2}
E_{\Pi_2}(A_1,\ldots ,A_q)=\sum_{\Pi_1\leq\Pi\leq\Pi_2}C_{\Pi_1,\Pi}(A_1,\ldots ,A_q)
\end{equation}

Let $q$ be an integer and $\sigma$ a permutation of $\Sy{q}$. 
For any $i\in [1,q]$ call $B_i$ the random variable
$B_i^d=U_{ii}\overline{U_{i\sigma(i)}}$, where $U_{ij}$ is the complex random variable on the
unitary group $U_d$ that was introduced in section \ref{ss-2.1}. For any subset $V\subset [1,q]$
we define the random variable 
\begin{equation}
B_V=\prod_{i\in V}B_i
\end{equation}
In particular, for $\Pi=\{V_1,\ldots ,V_k\}$ a partition of $[1,q]$ satisfying
$\Pi\geq \Pi_{\sigma}$, let
\begin{equation}\label{cum1}
E_{\Pi}(\sigma, d)=\prod_{i=1}^kE(B_{V_i})
\end{equation}
and 
\begin{equation*}
\begin{split}
C_{\Pi}(\sigma ,d)=\sum_{\Pi\leq\Pi_2} E_{\Pi} (\sigma, d)\Moeb(\Pi,\Pi_2) \\
C_{\Pi_1,\Pi_2}(\sigma, d)=\sum_{\Pi_1\leq\Pi\leq\Pi_2} E_{\Pi} (\sigma, d)\Moeb(\Pi,\Pi_2)\\
\end{split}
\end{equation*}
the classical (resp. relative) cumulants related to $E_{\Pi}(\sigma, d)$.

\begin{Thm}\label{DLexplicite}
Let $\sigma$ be an element of $\Sy{q}$.
Let $\Pi$ be a partition of $[1,q]$, such that $\Pi_{\sigma}\leq \Pi$.
Let $\tau$ be a permutation such that $\Pi_{\tau}=\Pi$. 
For integers $k,l$, let $\gamma_{\sigma,\Pi,k,l}$ the number of solutions of
$\sigma\sigma_1\ldots\sigma_k=e$ such that for all $i$, one has $\sigma_i\neq e$,  
$\sum_{i=1}^k|\sigma_i|=l$ and the group generated by $\tau,\sigma_1,\ldots,
\sigma_k$ acts transitively on $[1,q]$. Let
\begin{equation}\label{cum2}
\gamma_{\sigma,\Pi,l}=\sum_{k= 1}^l(-1)^k \gamma_{\sigma,\Pi,k,l}
\end{equation}
Let $d$ be an integer such that $d\geq q$, then
\begin{equation}\label{eqDLexplicite}
C_{\Pi,1_q}(\sigma ,d)=\sum_{l\geq 1} \gamma_{\sigma,\Pi,l} d^{-q-l}
\end{equation}
\end{Thm}

In order to prove theorem \ref{DLexplicite} we need some elementary
technical lemmas. We first derive an important corollary.
For a partition $\Pi$ of $[1,q]$, call $C(\Pi)$ the number of its blocks.

\begin{Cor}\label{decay}
Let $d,q$ be integers, $\sigma\in\Sy{q}$ be a permutation and
$\Pi$ be a partition of $[1,q]$ such that $d\geq q$. Then
$C_{\Pi_1,\Pi_2}(\sigma ,d)$ is a rational fraction of order
at most $-q-|\sigma|+2(C(\Pi_2)-C(\Pi_1))$.
\end{Cor}

We first prove corollary \ref{decay}.

\begin{demo}
From theorem \ref{DLexplicite} it is enough to prove that if 
$\tau\in\Sy{q}$ is such that $\Pi_{\tau}=\Pi$ and 
$\sigma_1,\ldots ,\sigma_k$ are permutations of $\Sy{q}$ such that
$\sigma\sigma_1\ldots\sigma_k=e$, and the group generated by
$\tau,\sigma_1,\ldots ,\sigma_k$ acts transitively on $[1,q]$,
then
$|\sigma_1 |+\ldots + |\sigma_k |\geq |\sigma|+2(C(\Pi)-1)$.

It is known (see e.g. \cite{MR2001g:05006}), that if $\sigma_1,\ldots ,\sigma_k$
act transitively on $[1,q]$ and satisfy $\sigma_1\ldots\sigma_k=\sigma$, then
$|\sigma_1|+\ldots +|\sigma_k |\geq 2q-2$.
In order to prove this, the difference is interpreted as the genus of
a given graph drawn on a two dimensional compact manifold.
This implies that if $\tau '$ is a permutation such that
$\Pi_{\tau '}=\vee_{i=1}^k\Pi_{\sigma_i}$, then
$|\sigma_1|+\ldots +|\sigma_k |\geq 2|\tau '|$.
Indeed, restricting the equation
$\sigma_1\ldots\sigma_k=\sigma$ to an orbit of $\tau '$ turns the action into
a transitive one, so the inequality of \cite{MR2001g:05006} holds.
Conclusion follows by summing on the orbits.

Thus it is enough to show that if 
the group generated by $\tau$ and $\tau '$ acts transitively on
$[1,q]$ then $|\tau '|+1\geq C(\Pi_{\tau})$.
Let $V_1,\ldots ,V_k$ be the blocks of $\Pi$, and 
$\tau_1\ldots\tau_{l}=\tau '$ be a cycle decomposition of $\tau '$.
This decomposition yields a graph on $V_1,\ldots ,V_k$ in the following
way~: if $\tau_i=(ab)$ with $a\in V_a$ and $b\in V_b$ then put
an edge between $V_a$ and $V_b$. 
This diagram has to be connected, so that $l\geq k-1$.
\end{demo}

\begin{Lem}\label{conv1}
Let $k_1,k_2$ be two integers satisfying $1\leq k_1 \leq k_2$. We have
\begin{equation}
\sum_{k\geq 1,V_1,V_2\subset[1,k],\, V_1\cup V_2=[1,k]\atop 
\card V_1=k_1, \,\card V_2=k_2 ,\, V_1\cup V_2=[1,k]} (-1)^k=(-1)^{k_1+k_2}
\end{equation}
\end{Lem}

\begin{demo}
Consider the formal power series
\begin{equation*}
A=(1+X)^{-1}(1+Y)^{-1}=\sum_{k_1,k_2\geq 0}A_{k_1,k_2}X^{k_1}Y^{k_2}
\in\mathbb{C}[[X,Y]]
\end{equation*} 
It is clear that $A_{k_1,k_2}=(-1)^{k_1+k_2}$. And if we rewrite 
$A$ as $(1+(X+Y+XY))^{-1}=\sum_k(-1)^k(X+Y+XY)^k$, then
\begin{equation*}
A_{k_1,k_2}=\sum_k\sum_{a_1,\ldots,a_k\in\{X,Y,XY\},\prod_ia_i=X^{k_1}Y^{k_2}}(-1)^k
\end{equation*}
To a sequence $(a_1,\ldots ,a_k)$ we can associate the pair $(V_1,V_2)$ of subsets of $[1,k]$
defined by $i\in\Pi_1$ iff $a_i=X$ or $XY$ and
$i\in V_2$ iff $a_i=Y$ or $XY$. We have $\card V_1=k_1$,
$\card V_2=k_2$ and $V_1\cup V_2=[1,k]$. This correspondence is clearly one to
one.
\end{demo}

\begin{Lem}\label{conv2}
Let $f$ and $g$ be functions $\mathbb{N}^*\rightarrow\mathbb{C}$, almost everywhere zero.
Define 
\begin{equation}\label{cconv}
S(f)=\sum_{k\geq 1} (-1)^k f(k)
\end{equation}
And define the operation
\begin{gather}\label{conv0}
h=f*g : \mathbb{N}^*\rightarrow\mathbb{C} \\
h(k) =  \sum_{\Pi_1,\Pi_2\subset [1,k],\Pi_1\cup\Pi_2=[1,k]}f(\card\Pi_1 )g(\card\Pi_2 )
\end{gather}
This operation is associative. Besides, we have for all $f,g$, 
$S(f*g)=S(f)S(g)$.
\end{Lem}

\begin{demo}
Associativity is elementary. For the multiplicativity, one just needs to remark
that $g(k_1)h(k_2)$ contributes to $(-1)^{k_1+k_2}$ in $S(f)S(g)$, and that it 
gives a contribution of
\begin{equation*}
\sum_{k\geq 1\Pi_1,\Pi_2\subset[1,k],|\Pi_1|=k_1,|\Pi_2|=k_2,
\Pi_1\cup\Pi_2=[1,k]} (-1)^k
\end{equation*}
in $S(h)$, so this is a consequence of the lemma \ref{conv1}.
\end{demo}

Let $\Pi=\{ V_1,\ldots ,V_j \}$ be a partition of $[1,q]$; $k$ and $l$ be integers and
consider a $k$ -uple $(\sigma_1,\ldots ,\sigma_k)$ such that
$\sigma\sigma_1\ldots \sigma_k=e$, for all $i$, $\Pi_{\sigma_i}\leq \Pi$ and
$\sigma_i\neq e$.
We call $A[\sigma ,k,l](\Pi)$ the number of such $k$ -uples. 
We also need the notation $A[\sigma ,k,l](V_i)=A[\sigma_{|V_i} ,k,l]$.
As usual we denote $A[\sigma ,l](\Pi)=S(A[\sigma , \cdot ,l](\Pi))$
and $A[\sigma ,l](V_i)=S(A[\sigma , \cdot ,l](V_i))$.

\begin{Lem}\label{mult}
Under the preceding notations, we have
\begin{equation}
A[\sigma ,l](\Pi)=\sum_{l_1,\ldots ,l_j,\sum_{r=1}^j l_r=l}\prod_{r=1}^j 
A[\sigma ,l_r](V_r)
\end{equation}
\end{Lem}

\begin{demo}
If $\Pi=\{ V_1,\ldots ,V_j \}$, and $l_1 ,\ldots ,l_j$ are integers
we call $A[\sigma ,k,[l_1,\ldots ,l_j]](\Pi)$ the number of solutions of
the equation $\sigma\sigma_1\ldots \sigma_k=e$ in $\Sy{q} - \{ e\}$ satisfying
for all $r\in [1,j]$
$\sum_{i=1}^k |\sigma_{i|V_r} |=l_r$ and for all $i\in [1,k]$, $[\sigma_i]\leq [\Pi]$.
We have 
\begin{equation*}
A[\sigma ,l](\Pi)=\sum_{l_1,\ldots ,l_j,\sum_{r=1}^j l_r=l}
A[\sigma ,[l_1,\ldots ,l_j]](\Pi)
\end{equation*}
so it is enough to prove that 
$$A[\sigma ,k,[l_1,\ldots ,l_j]](\Pi)=\prod_{r=1}^j A[\sigma ,l_r](V_r)$$
For this, introduce the functions $f_r$ for $1\leq r\leq j$ defined as
$k\geq 1$ by $f_r(k)=A[\sigma ,k,l_r](V_r)$.
Define
$f$ by $f(k)=A[\sigma ,k,[l_1,\ldots ,l_j]](\Pi)$.

Considering the convolution that we introduced in equation \ref{conv0}
we have $f=f_1*\ldots *f_k$ so by the lemma \ref{conv2} we get
$S(f)=S(f_1)\ldots S(f_k)$.
\end{demo}

\begin{Lem}\label{bof}
Let $\Pi$ be a partition of $[1,k]$ then $E_{\Pi}(B)[l]=A[\sigma, l](\Pi)$
\end{Lem}

\begin{demo}
By theorem \ref{combWe},
we have $E_{V_i}(B)[l]=A[\sigma, l](V_i)$ so 
$E_{\Pi}(d,\sigma )[l]=\sum_{l_1,\ldots ,l_j,\sum_{r=1}^j l_r=l}\prod_{r=1}^j 
A[\sigma ,l_r](V_r)$, which equals
$A[\sigma, l](\Pi)$ by lemma \ref{mult}.
\end{demo}

Now we can go through the proof of theorem \ref{DLexplicite}.

\begin{demo}
We have by definition
$C_{\Pi,1_q}(\sigma ,d)[l]=\sum_{\Pi'\geq \Pi}E_B(\Pi)[l]\Moeb (\Pi',1_q)$. 
By lemma \ref{bof} this is
$\sum_{\Pi'\geq \Pi}A[\sigma, l](\Pi')\Moeb (\Pi',1_q)$.

Let $\Pi$ be a partition such that $\Pi_{\sigma}\leq \Pi$, and
let $A[\sigma, k, l][\Pi]$ be the number of $k$ -uples
of permutations $(\sigma_1, \ldots ,\sigma_k)$ such that
$\sigma_i\neq e$ for all $i$ ; $\sigma\sigma_1\ldots\sigma_k=e$ and
$\sum_i{|\sigma_i|}=l$ and $\vee_i [\sigma_i]=\Pi$.
Define as usual $A[\sigma, l][\Pi]=S(A[\sigma, \cdot , l][\Pi])$.
We have the obvious relation between
$A[\sigma, l][\Pi]$ and $A[\sigma, l](\Pi)$
\begin{equation}
A[\sigma, l](\Pi)=\sum_{\Pi'}1_{\Pi\geq\Pi'}A[\sigma, l][\Pi']
\end{equation}

By proposition \ref{rota2} this implies that
$C_{\Pi,1_q}(\sigma, d)[l]=A[\sigma, l][\Pi]$, which is the expected result.
\end{demo}

\subsection{A first order expansion for $C_{\Pi,1_q}(d,\sigma)$}

So far we have obtained upper bounds for the degrees of cumulants of unitary
polynomial integrals. However, except in some very elementary cases, it
is far from obvious that they are optimal. 
In this section we show that our bounds are optimal and perform some explicit
computations.

Let $\Moeb : \Sy{q}\rightarrow\mathbb{C}$ be the 
central function defined by
\begin{equation}\label{defsp}
\Moeb (\sigma ):=\prod_{i=1}^{k}c_{|C_i|}\cdot (-1)^{|C_i|}
\end{equation}
where $\sigma\in\mathcal{S}_q$ is a 
permutation whose cycle decomposition is $\sigma=C_1\cdots C_k$ and
$c_n=(2n)!/(n!(n+1)!)$. The integer $c_n$ is the {\it $n^{th}$ Catalan number \it} 
and satisfies the inductive relation $c_n=\sum_{i=0}^{n-1}c_i\cdot c_{n-i-1}$.
Its first values are 
\begin{equation*}
c_o=1, c_1=1, c_2=2, c_3=5, c_4=14 c_5=42, c_6=132, c_7=429, c_8=1430
\end{equation*}
\begin{rem}
This function
is related to Speicher's function $\Moeb$ defined in \cite{MR95h:05012}
in the sense that a non crossing partition $\Pi$
(that will be defined at section \ref{ss4-4})
of $[1,q]$ with blocks of length $|C_1|,\cdots ,|C_k|$ satisfies 
$\Moeb (\Pi)=\Moeb (\sigma)$.
\end{rem}

\begin{Thm}\label{Schaeffer}
\begin{itemize}
\item (i)
Let $\sigma$ be a permutation of $[1,q]$ such that 
in its cycle product decomposition
it has $d_i$ cycles of length $i-1$. Then
\begin{equation}\label{conj}
\gamma_{\sigma, \Pi_{\sigma}, 3q-2-|\sigma|}=(-1)^{|\sigma |}
\frac{2^{q-|\sigma|}(3q-3-|\sigma|)!}{(2q)!}
\prod_{i=1}^q \left(\frac{(2i-1)!}{(i-1)!^2} \right)^{d_i}
\end{equation}
where $\gamma$ was defined in theorem \ref{DLexplicite}, equation \ref{cum2}.
\item (ii)
In particular,
\begin{equation}\label{eqmoeb}
\lim_{d\rightarrow\infty}d^{q+|\sigma |}\We (d,\sigma)=\Moeb(\sigma)
\end{equation}
\item (iii)
Let $\Pi'$ be a partition having blocks $V_1,\ldots ,V_k$
of length $q_1,\ldots ,q_k$ such that $\Pi_{\sigma}\leq\Pi$ 
and define $\sigma_i=\sigma_{|V_i}$.
Call 
\begin{equation}\label{mmyst}
g_{\sigma, \Pi}
=\sum_{\Pi'\in\mathcal{P}_q, \Pi'\geq \Pi_{\sigma}, \Pi'\vee\Pi=1_q
\atop C(\Pi_{\sigma})-C(\Pi')=C(\Pi)-1}
\prod_{i=1}^k \frac{(3q_i-3-|\sigma_i|)!}{(2q_i)!}
\end{equation}
we have
\begin{equation}\label{mystere3}
\gamma_{\sigma,\Pi, q+|\sigma|+2(C(\Pi)-1)}=
g_{\sigma, \Pi}
\prod_{i=1}^q \left(\frac{(2i-1)!}{(i-1)!^2} \right)^{d_i}
(-1)^{|\sigma |}2^{q-|\sigma |}
\end{equation}
Note that this it is not zero.
\end{itemize}
\end{Thm}

\begin{demo}
Formula \ref{mystere3} is a consequence of \ref{conj} and \ref{eqmoeb}
together with lemma \ref{mult}. The fact that it is non zero comes from
the fact that all summands of the right hand side of \ref{mmyst}
have the same sign and are non zero.

So we focus on proving \ref{conj} and \ref{eqmoeb}. 
In \cite{MR2001g:05006}, the following result is proved:
the number of $k$ -uples $(\sigma_1,\ldots ,\sigma_k)$
of permutations of $\Sy{q}$ such that
$\sigma_1\ldots\sigma_k\sigma=e$, the group generated by
$\sigma_1,\ldots ,\sigma_k$ acts transitively on $[1,q]$ 
and $|\sigma|+|\sigma_1|+\ldots +|\sigma_k|=2q-2$, is
\begin{equation*}
\tilde{A}[\sigma ,k]=k\frac{(qk-q-1)!}{(qk-2q+|\sigma|+2)!}\prod_{i\geq 1} \left[ i 
\left(_i^{ki-1} \right)\right]^{d_i}
\end{equation*}
where $d_i$ denotes the number of cycles with $i$ elements
of $\sigma$. 

This set allows the $\sigma_i$'s to be identity. An application of the 
exclusion-inclusion principle shows that: the 
number of $k$ -uples $(\sigma_1,\ldots ,\sigma_k)$
of permutations of $\Sy{q}$ {\it different from identity \it} such that
$\sigma_1\ldots\sigma_k\sigma=e$, the group generated by
$\sigma_1,\ldots ,\sigma_k$ acts transitively on $[1,q]$ 
and $|\sigma|+|\sigma_1|+\ldots +|\sigma_k|=2q-2$, is
\begin{equation}
\gamma_{\sigma,\Pi_{\sigma},k,2q-2-|\sigma|}
=\sum_{l=1}^{2q-2-|\sigma|} {2q-2-|\sigma| \choose l}
\tilde{A}[\sigma ,l](-1)^l
\end{equation} 
We need to estimate $\gamma_{\sigma,\Pi_{\sigma},2q-2-|\sigma|}
=S(\gamma_{\sigma,\Pi_{\sigma},k,2q-2-|\sigma|})$.
The fact that for $k'\geq k$ one has
$\sum_{n=k}^{k'}{n \choose k}={k'+1 \choose k+1}$ implies that

\begin{itemize}
\item (i)
If $|\sigma|<q-1$ then
\begin{equation}\label{mystere1}
\begin{split}
\gamma_{\sigma,\Pi_{\sigma} ,2q-2-|\sigma|}=\\
\sum_{k=2}^{2q-2-|\sigma|}(-1)^k { 2q-1-|\sigma| \choose k+1}
k\frac{(qk-q-1)!}{(qk-2q+|\sigma|+2)!}\prod_{i\geq 1} \left[ i 
{ki-1 \choose i} \right]^{d_i}\\
\end{split}
\end{equation}
\item (ii)
if $|\sigma|=q-1$ then 
\begin{equation}\label{mystere2}
\gamma_{\sigma,\Pi_{\sigma}, 2q-2-|\sigma|}=
\sum_{k=2}^q(-1)^{q-k}\frac{(qk-q)!}{k!(q-k)!(qk-2q+1)!}
\end{equation}
\end{itemize}
Now we need to prove that for $\sigma\in\Sy{q}$ having $a_i$ cycles with $i$ 
elements, one has
\begin{equation}
\gamma_{\sigma, \Pi_{\sigma}, 3q-3-|\sigma|}=\frac{2^q(3q-2-|\sigma|)!}{(2q)!}
\prod_{i=1}^q \left(\frac{(2i-1)!}{(i-1)!^2(-2)^{i-1}} \right)^{d_i}
\end{equation}
We recall that for $0\leq r < n$ we have
\begin{equation}\label{adj}
\sum_{k=0}^n{n \choose k}k^r(-1)^k=0
\end{equation}
Indeed, the holomorphic function $z\rightarrow (1-\exp z)^n$ has a zero
of order $n$ at $0$ an the left hand side of equation \ref{adj} is its $r$-th 
derivative at zero. 
We first handle equation \ref{mystere1}. The expression
\begin{equation*}
k\frac{(qk-q-1)!}{(qk-2q+|\sigma|+2)!}\prod_{i\geq 1} \left[ i 
{ki-1 \choose i} \right]^{d_i}
\end{equation*}
is a polynomial in $k$, of
degree at most $2q-2-|\sigma|$. It is easily checked that 
$P(1)=P(0)=1$. 
So according to equation \ref{adj}.
\begin{equation*}
\sum_{k=2}^{2q-2-|\sigma|}(-1)^k { 2q-1-|\sigma| \choose k+1}
k\frac{(qk-q-1)!}{(qk-2q+|\sigma|+2)!}\prod_{i\geq 1} \left[ i 
{ki-1 \choose i} \right]^{d_i}=-P(-1)
\end{equation*}

But
\begin{equation}
-P(-1)=\frac{2^q(3q-3-|\sigma|)!}{(2q)!}
\prod_{i=1}^q \left(\frac{(2i-1)!}{(i-1)!^2(-2)^{i-1}} \right)^{d_i}
\end{equation}
This proves \ref{conj} in the case of \ref{mystere1}.
The case of \ref{mystere2} is done in the same way.
\end{demo}

\section{An application to free probability theory}\label{s-3}

In this section we first recall basic definitions about asymptotic freeness.
The main results of this section are Theorems
\ref{libre} and \ref{proba}. In order to prove
\ref{proba} we introduce material
that will be used as well in Section \ref{s-4}.

\subsection{Asymptotic freeness}\label{ss-2.1}

Our definition of a {\it non commutative probability space \it} is the
following: it is an algebra with unit endowed with a faithful tracial 
state $\phi$. In particular we do not make any assumption about
the linear form $\phi$, except that $\phi (ab)=\phi (ba)$ and $\phi (1)= 1$.
We do not need the $*$ -structure
of the matrix algebras (and a fortiori faithfulness or positivity assumptions).
We denote such a space by $(A, \phi )$. An element of this space will be called
a (non-commutative) random variable. 

Let $A_1, \cdots ,A_k$ be subalgebras of $A$ having the same unit as $A$.
They are said to be {\it free \it} iff
for all $a_i\in A_{j_i}$ ($j\in[1,l]$) 
such that $\phi(a_i)=0$, one has  $\phi(a_1\cdots a_l)=0$ as soon as
$j_1\neq j_2$, $j_2\neq j_3,\cdots ,j_{l-1}\neq j_l$.
Random variables are said to be free iff the 
unital subalgebras that they generate are free.

Let $(a_1,\cdots ,a_k)$ be a $k$ -uple of random variables and let
$\mathbb{C}\langle X_1 , \cdots , X_k \rangle$ be the
free algebra of non commutative polynomials on $\mathbb{C}$ generated by
the $k$ indeterminates $X_1, \cdots ,X_k$. 
The {\it joint distribution\it} of the family $a_i$ is the linear form
$\mu_{(a_1,\cdots ,a_k)} : 
\mathbb{C}\langle X_1, \cdots ,X_k \rangle
\rightarrow \mathbb{C}$
defined in the obvious sense. 

Given a $k$ -uple $(a_1,\cdots ,a_k)$ of free 
random variables and given each law $\mu_{a_i}$, the joint law
$\mu_{(a_1,\cdots ,a_k)}$ is uniquely determined by induction with the
$\mu_{a_i}$'s.
We shall say that a family $(a_1^d,\cdots ,a_k^d)_d$ of $k$ -uples of random
variables {\it converges in law \it} towards $(a_1,\cdots ,a_k)$
iff for all $P\in \mathbb{C}\langle X_1, \cdots ,X_k \rangle$, 
$\mu_{(a_1^d,\cdots ,a_k^d)}(P)$ converges towards
$\mu_{(a_1,\cdots ,a_k)}(P)$ as $d\rightarrow\infty$. 

This gives the following obvious sense to {\it asymptotic freeness\it}:
a sequence of families 
$(a_1^d,\cdots ,a_k^d)_d$ is asymptotically free as $d\rightarrow\infty$
iff it converges in law towards a free random variable. 

By now we are able to state the main result of this section. Note that it 
was originally contained in \cite{MR94c:46133} under stronger hypotheses. 
Feng Xu \cite{MR99f:81185} obtains an analogous result
using geometric methods.

\begin{Thm}\label{libre}
Let $U_{1},\cdots ,U_{k}, \cdots$ be a collection of independent 
Haar distributed random matrices of $\M{d}$ and $( W^d_i )_{i\in I}$ be a 
set of constant matrices of $\M{d}$ 
admitting a joint limit distribution for large $d$.
Then the family $( ( U_{1}, U_1^* ) ,\cdots , 
( U_{k}, U_k^* ), \cdots , (W_i))$ admits a limit distribution, and
is asymptotically free.
\end{Thm}

Note that this statement holds under
the very weak hypothesis that the joint law of the family $W$ admits a
weak limit, i.e. it does not make any assumption of boundedness of the
elements of $W$ as $d\rightarrow\infty$.
In other words we do not have to consider asymptotic $*$ -freeness,  
our method works in the framework of asymptotic algebraic  
freeness. This allows us to escape from the machinery of 
functional calculus and weaken the hypotheses of \cite{MR2000d:46080}.
Besides, we do not have to restrict to diagonal elements or 
self adjoint elements for the $(W_i)_i$ as in the previous proofs.
In particular they might have no $*$ -asymptotic limit.

We split the proof of this theorem into two parts. The first one consists in
showing that there exists an asymptotic joint law for the families
of random variables. The second step is to show that these families are
asymptotically free. 

Let $W_1,\cdots ,W_n$ be matrices of $\M{d}$ and $\alpha$ a permutation of
$[1,n]$. If the cycle decomposition of $\alpha$ is
$(a_{11}\cdots a_{1i_1})\cdots (a_{k1}\cdots a_{ki_k})$, then
we define
\begin{equation}\label{defW}
\langle W_1 \cdots W_n \rangle_{\alpha}:=
\tr (W_{a_{11}} \cdots 
W_{a_{1i_i}})\cdots
\tr (W_{a_{k1}} \cdots 
W_{a_{ki_k}} )
\end{equation}

These notations are well defined because we consider the tracial state 
$\tr $.
For these notations to be well defined, it
is essential that we consider a tracial non commutative probability space.

For an integer $n$, let $\xi$ be a permutation of $\Sy{n}$ such that there exists 
disjoint subsets $V_1$
and $V_2$ with $q$ elements of $[1,n]$ such that $\xi(V_1)=V_2$, $\xi(V_2)=V_1$ and
$\xi$ stabilizes pointwise $(V_1\cup V_2)^c$.
We can check promptly that $\xi^2$ stabilizes both $V$ and $V^c$, and that
the restrictions of $\xi^2$ to $V_1$ and to $V_2$ are conjugate. 
This allows to define 
\begin{equation*}
\widetilde{\We}(\xi)=\We(d,\xi^2_{|V})
\end{equation*}
Besides  it is routine to check that $\We(\xi)=0(d^{-|\xi|})$.
From this we can state

\begin{Prop}\label{eval2}
Let $U_1,\ldots ,U_k$ be $k$ independent Haar distributed random matrices.
Let $n=2q$ be an integer and for $i\in [1,n]$ let 
$\widetilde{W}_i$ be a random matrix of the
form $W_iU_{j_i}^{\varepsilon_i}$ with $j_i\in [1,k]$ and $\varepsilon_i\in\{-1,1\}$.
For $j\in [1,k]$, let $\mathcal{T}_j$ be the subset of $\Sy{n}$ of permutations $\xi$
such that if $j_i=k$ then $j_{\xi(i)}=k$ and 
$\varepsilon_{\xi(i)}=-\varepsilon_i$ and if $j_i\neq k$ then $\xi(i)=i$.
Let $\alpha$ be a permutation of $[1,n]$. Then 
\begin{equation}\label{eqeval2}
E(\langle \widetilde{W}_1\ldots \widetilde{W}_n\rangle_{\alpha})=
\sum_{\xi_1\in\mathcal{T}_1,\ldots ,\xi_k\in\mathcal{T}_k}
\langle W_1\ldots W_n \rangle_{\alpha\xi_1\ldots\xi_k}
d^{|\alpha |-|\alpha\xi_1\ldots\xi_k |}
\prod_{i=1}^k\widetilde{\We} (\xi_i) 
\end{equation}
\end{Prop}

\begin{demo}
This is done by induction on $k$. For $k=1$ this is a standard combinatorial 
computation in view of Theorem \ref{thWe}.
For the general $k$ case, it is enough to apply Fubini's theorem and
remark that the elements of 
$\mathcal{T}_i$ and $\mathcal{T}_j$ commute.
\end{demo}

Each summand of equation \ref{eqeval2} is of order $0(1)$ because by 
the triangle inequality $|\alpha |-|\alpha\xi_1\ldots\xi_k|-
|\xi_1|-\ldots -|\xi_k|\leq 0$. This proves the existence of a joint distribution.

\begin{Prop}
\begin{itemize}
\item (i)
In equation \ref{eqeval2}, if for a given $\xi_1\in\mathcal{T}_1,\ldots ,
\xi_k\in\mathcal{T}_k$,
one has $|\alpha|-|\alpha\xi_1\ldots\xi_k|-|\xi_1|-\ldots -|\xi_k|=0$ 
then $\alpha\xi_1\ldots\xi_k$ admits at least two fixed points. 
Besides if $j$ is a fixed point, it can not be such that
$\varepsilon_{j-1}=\varepsilon_j$ and $i_{j-1}=i_j$.
\item (ii)
In particular the family $((W),(U_1, U_1^*), \cdots ,(U_k, U_k^*))$ 
is asymptotically free.
\end{itemize}
\end{Prop}

\begin{demo}

The fact that the first point implies the second one is as for the proof of
\ref{eqeval2}, a straightforward consequence of the definition of freeness. 
Let us consider the situation
$|\alpha|-|\alpha\xi_1\ldots\xi_k|-|\xi_1|-\ldots -|\xi_k|=0$.
We have 
$|\alpha\xi_1\ldots\xi_k| \leq |\alpha | -q \leq q-1$. 
Consequently $\phi$ moves at most
$2(q-1)=n-2$ elements, hence has at least two fixed points. Besides, according
to the definition of $\phi$ and of $\mathcal{T}_i$, 
the permutation $\alpha\xi_1\ldots\xi_k$ has to move the point $i$
if $\varepsilon_{j-1}=\varepsilon_j$ and $i_{j-1}=i_j$.
\end{demo}

Since the proposition holds for all $k$, it holds for an arbitrary family
of independent random variables, therefore
this proposition proves the theorem \ref{libre} in full generality.

The following corollary can be found in
\cite{MR2000d:46080} and \cite{MR99f:81185} and it is a consequence from
Theorem \ref{libre}.

\begin{Cor}
Let $W$ be a family of constant matrices with a limiting distribution
and  $U_1,\cdots ,U_n$ be random unitary matrices. Then the variables
$W, U_1 W U^*_1, \cdots ,  U_k W U^*_k$ are asymptotically free.
\end{Cor}

\begin{demo}
This is a straightforward consequence of the theorem \ref{libre} and of the 
definition of asymptotic freeness.
\end{demo}

\begin{rem}
The results we obtained are the same if
one replaces $\mathbb{U}_d$ by $SU_d$.
Indeed, a careful reading of the proof of theorem \ref{thWe} shows that
it also holds for $SU_d$. Another way to see it is to note that
the multiplication map
$SU_d\times \mathbb{U}_1\rightarrow \mathbb{U}_d$ is both a group morphism
and a probability space morphism (for these groups endowed with
their respective normalized Haar measure).
\end{rem}

\begin{rem}
A slight modification of the argument of the proof of Theorem \ref{libre}
shows that the result holds as well provided that for all $q\in\mathbb{N}$
and for all $\sigma\in\Sy{q}$,
\begin{equation*}
E(\langle W \rangle_{\sigma})=\prod_{c \,\, {\rm cycle \,\, of \rm} \,\, \sigma}
E(\langle W \rangle_{c})+o(1)
\end{equation*}
as $d\rightarrow\infty$. In particular, this yields
a new proof of asymptotic freedom of independent $GUE$'s (albeit much more
complicated). 
\end{rem}

\subsection{Refinements and corollaries of theorem \ref{libre}}\label{ss-2.3}

Our methods allow to derive an interesting consequence of theorem \ref{libre}:

\begin{Thm}\label{proba}
Let $W$ be a family of matrices admitting a limit law and $U_{1}, \cdots ,U_{k}$ 
be unitary independant random variables. Let $w,u_{1},\cdots ,u_{k}$ be 
non-commutative random variables whose law is the limit joint law of 
$W,U_{1},\cdots ,U_{k}$.
If $\varepsilon > 0$, then one has
\begin{equation} \label{eq4}
P( | \langle W_1 U_{i_1}^{\varepsilon_1} \cdots
W_n U_{i_n}^{\varepsilon_n} \rangle - \langle w_1 u_{i_1}^{\varepsilon_1}\cdots
w_n u_{i_n}^{\varepsilon_n} \rangle | \geq \varepsilon)= O(d^{-2})
\end{equation}
In particular, the random variable
$\langle W_1 U_{i_1}^{\varepsilon_1} \cdots W_n U_{i_n}^{\varepsilon_n} \rangle$
converges in probability  towards
$\langle w_1 u_{i_1}^{\varepsilon_1}\cdots
w_n u_{i_n}^{\varepsilon_n} \rangle$
\end{Thm}

In order to prove this we need to estimate the expectation of the
product of the traces of random variables in terms of the product of the
expectations of the traces of the random variables. 
Corollary \ref{facto} of the following proposition does the job, 
but we shall also use proposition \ref{gcumu} in section \ref{s-4}.
We define the relative cumulants 
$C_{\Pi_1,\Pi_2}(\langle \widetilde{W}_1\ldots \widetilde{W}_n\rangle_{\alpha})$
by modifying the definition of equation \ref{cumrel} in the obvious
way.

\begin{Prop}\label{gcumu}
Take the same notations as in lemma \ref{eval2}. Then
\begin{equation}
C_{\Pi_1,\Pi_2}(\langle \widetilde{W}_1\ldots \widetilde{W}_n\rangle_{\alpha})
=0(d^{2(C(\Pi_2)-C(\Pi_1))})
\end{equation}
\end{Prop}

\begin{demo}
We have, according to lemma \ref{eval2}, 
\begin{equation*}
C_{\Pi_1,\Pi_2}(\langle \widetilde{W}_1\ldots \widetilde{W}_n\rangle_{\alpha})=
\sum_{\xi_1\in\mathcal{T}_1,\ldots ,\xi_k\in\mathcal{T}_k}
\langle W_1\ldots W_n \rangle_{\alpha\xi_1\ldots\xi_k}
C_{\Pi,\Pi_2}(\prod_{i=1}^k\widetilde{\We} (\xi_i)) 
\end{equation*}
with $\Pi=\Pi_1\vee[\xi_1]\vee\ldots\vee[\xi_k]$. 
But it is straightforward to check that
\begin{equation*} 
C_{\Pi,\Pi_2}(\prod_{i=1}^k\widetilde{\We} (\xi_i))=O(d^{-|\xi_1\ldots \xi_k|+2(C(\Pi_2)-C(\Pi))})
\end{equation*}
and for $\xi_1,\ldots ,\xi_k$ such that $\Pi=\Pi_1\vee\Pi_{\xi_1}\vee\ldots\vee\Pi_{\xi_k}$ 
one has
\begin{equation*}
|\alpha|-|\alpha\xi_1\ldots\xi_k|\leq 2(C(\Pi)-C(\Pi_1))+|\xi_1|+\ldots +|\xi_k| 
\end{equation*}
\end{demo}

This proposition implies the following corollary:

\begin{Cor}\label{facto}
Let $W_i$ and $U$ be as in the equation \ref{defW}.
The following holds
\begin{equation*}
\begin{split}
E(\langle W_1 U^{\varepsilon_1} \cdots W_n U^{\varepsilon_n}
\rangle_{\alpha_1}
\langle W_{n+1} U_{i_{n+1}}^{\varepsilon_{n+1}} \cdots W_m U_{i_n}^{\varepsilon_m}
\rangle_{\alpha_2})=\\
E(\langle W_1 U_{i_1}^{\varepsilon_1} \cdots W_n U_{i_n}^{\varepsilon_n}
\rangle_{\alpha_1}) E(
\langle W_{n+1} U_{i_{n+1}}^{\varepsilon_{n+1}} \cdots W_m U_{i_m}^{\varepsilon_m}
\rangle_{\alpha_2}) + O(d^{-2})\\
\end{split}
\end{equation*}
\end{Cor}

Now we can proceed to the proof of Theorem \ref{proba}

\begin{demo}
We prove a slightly better result, namely
\begin{equation*}
E( | \langle W_1 U_{i_1}^{\varepsilon_1} \cdots
W_n U_{i_n}^{\varepsilon_n} \rangle - \langle w_1 u_{i_1}^{\varepsilon_1}\cdots
w_n u_{i_n}^{\varepsilon_n} \rangle |^2)= O(d^{-2})
\end{equation*}
Our result will follow by Tshebyshev inequality. Developing the above square
yields that it is enough to show that 
\begin{equation*}
E( | \langle W_1 U_{i_1}^{\varepsilon_1} \cdots
W_n U_{i_n}^{\varepsilon_n} \rangle|^2) - |E( \langle W_1 U_{i_1}^{\varepsilon_1}\cdots
W_n U_{i_n}^{\varepsilon_n} \rangle)|^2 =O(d^{-2})
\end{equation*}
This is an immediate consequence of the lemma \ref{facto} on the factorization of the
expectation.
\end{demo}

\begin{rem}
\begin{itemize}
\item
In corollary \ref{proba}, for the convergence in
probability to hold in equation \ref{eq4}, it is enough for the family $\{ W_i \}$ to 
have joint $2n^{th}$ moments that are bounded uniformly in $d$.
\item
If we assume furthermore that 
the random variables $(U_i^d\in\M{d})_{d\geq 1}$ are defined on the same large 
probability space $\Omega$,  then the convergence
holds almost surely. This is just Borel-Cantelli's lemma and the fact that 
$\sum d^{-2}$ converges.
\end{itemize}
\end{rem}

\section{Large dimension behaviour of the Itzykson-Zuber integral}\label{s-4}

We start with two theorems which were one of our initial motivations for performing
explicit computations on functions of the kind $d\rightarrow 
C_{\Pi_1,\Pi_2}(d,\sigma)$. 

\begin{Thm}\label{universal}
Let $W$ be a family of matrices admiting a limit joint distribution. 
Let $U_1, \ldots ,U_k$ be independent Haar distributed
unitary matrices. 
Let $(P_{i,j})_{1\leq i \leq k , 1\leq j \leq k}$ 
and
$(Q_{i,j})_{1\leq i \leq k , 1\leq j \leq k}$ be two families of noncommutative polynomials
in $U_1,U_1^*,\ldots ,U_k,U_k^*$ and $W$.
Let $A_d$ be the variable $\sum_{i=1}^k\prod_{j=1}^k\tr P_{i,j}(U,U^*,W)$ and
$B_d$ the variable $\sum_{i=1}^k\prod_{j=1}^k\tr Q_{i,j}(U,U^*,W)$
\begin{itemize}
\item (i)
For each $d$, the analytic function
\begin{equation*}
z\rightarrow d^{-2}\log E \exp (zd^2A_d)
=\sum_{n\geq 1}a_n^d z^n
\end{equation*}
is such that for all $q$, $\lim_d a_q^d$ exists and is finite. 
It depends only on the limit distribution of $W$ and on the polynomials $P_{i,j}$.
\item (ii)
For each $d$, the analytic function
\begin{equation*}
z\rightarrow \frac{E \exp (zB_d+zd^2A_d)}{E \exp (zd^2A_d)}=1+\sum_{n\geq 1}b_n^d z^n
\end{equation*}
is such that for all $q$, $\lim_d b_q^d$ exists. It
depends only on the limit distribution of $W$ and on the polynomials $P_{i,j}$ and $Q_{i,j}$.
\end{itemize}
\end{Thm}

\begin{demo} 
For the first point, remark that if $q$ is greater than $1$ then 
$q!a_q^d=C_q(d^2A)d^{-2}$. By Proposition \ref{gcumu}
this is asymptotically bounded.
For the second point we show that equivalently, defining
\begin{equation*}
\log (1+\sum_{q\geq 1}b_q^d z^q )=\sum_{q\geq 1}b_q^{'d} z^q
\end{equation*}
the coefficients $b_q^{'d}$ are such that for all $q$ greater than
$1$, $\lim_db_q^{'d}$ exists, is bounded. But 
\begin{equation*}
q!b_q^{'d}=C_q(B+d^2A)-C_q(d^2A)
\end{equation*}
Developing this by multilinearity and applying Proposition \ref{gcumu}
shows that $q!b_q^{'d}=qC_q(B,d^2A,\ldots ,d^2A)+o(1)$ and
$\lim_d C_q(B,d^2A,\ldots ,d^2A)$ itself exists, so
this concludes our proof.
\end{demo}

\begin{rem}
Heuristically, point (ii) of theorem \ref{universal} is a way to
give a meaning to $E(\exp (B+d^2A))/E(\exp d^2A)$ in the large $d$ limit.
However, it is important to note that conclusion of (ii) does not
hold any more if one considers, for example 
$z\rightarrow E(B\exp zd^2A)/E(\exp zd^2A)$.
\end{rem}

In the remainder, we set $(X_d)_{d\geq 1}$, $(Y_d)_{d\geq 1}$ 
to be two families of matrices
of $\M{d}$ and consider them as non-commutative
variables of the non commutative probability space $(\M{d}, \tr )$.
Let $U$ be a matricial random variable with values in the unitary group
$\mathbb{U}_d$ and with law the left and right invariant Haar measure.
Consider the commutative random variable $A_d=\tr (X_dUY_dU^*)$.
We denote by $A$ the family $A=(A_d)_{d\geq 1}$.
The {\it Itzykson-Zuber integral \it} (or IZ integral)
is the following analytic function in a neighbourhood of $0$:

\begin{equation} \label{iz1}
IZ_{d,X,Y}: z\rightarrow E(\exp (d^2 z \tr(A_d)))
\end{equation}

Let $(x_i)$ and $(y_j)$ be the eigenvalues of $X$ and $Y$ 
and $\Delta(X)=\prod_{1\leq i<j\leq d}(x_j-x_i)$ 
(resp. $\Delta(Y)=\prod_{1\leq i<j \leq d}(y_j-y_i)$) be the Vandermonde determinant
associated to the roots $(x_i)$ (resp. $(y_j)$).
The IZ integral first appeared in relation to matrix models and two-dimensional
quantum gravity (\cite{MR81a:81068}). It can be computed explicitely by 
the Harish-Chandra formula provided that $x_i=x_j$ iff $i=j$:

\begin{equation} \label{hc}
E(\exp (d^2\tr (A)))=
\frac{\det (e^{d\cdot x_iy_j})_{1\leq i,j\leq d}}{\Delta(X)\Delta(Y)}
\end{equation}

In this section we investigate the function
\begin{equation} \label{iz2}
F_{d,X,Y}: z\rightarrow d^{-2}\log E(\exp (d^2 z \tr(A_d)))
\end{equation}
In \cite{MR94m:81104}, 
Matytsin identifies the limit of $F_{d,X,Y}(1)$ with the solution of
some $PDE$ with boundary conditions,
under the assumption that the distributions of $X$ and $Y$ admit a smooth
limit.
The existence of a large $d$ limit remains a non-obvious mathematical
problem. It has been solved very recently by A. Guionnet and O. 
Zeitouni in \cite{MR1883414} under the assumption that $X$ and $Y$ admits a 
limit bounded distribution and are Hermitian.
However, here we investigate the limit of $F$ 
by focusing on the study of the coefficients of its series in $0$:
\begin{equation*}
d^{-2}C_q(d^2A)=\frac{\partial ^q}{\partial z^q} F_{d,X,Y}(0)
\end{equation*}
In the remainder we define $\langle X \rangle_{\sigma}$ to be
$\langle X^q \rangle_{\sigma}$, which was defined in Equation
\ref{defW}.

\begin{Thm}\label{Fborne} 
We have $\lim_d d^{-2}C_q(d^2A_d)=$
\begin{equation}
\sum_{\sigma,\tau\in \mathcal{S}_q \atop 
|\tau|+|\sigma|+|\tau\sigma^{-1}|=2q-2C(\Pi_{\tau}\vee \Pi_{\sigma})}
\langle X\rangle_{\sigma}\langle Y \rangle_{\tau}
\gamma_{\tau\sigma^{-1},\Pi_{\tau}\vee\Pi_{\sigma}, 
q+|\tau\sigma^{-1}|+2(C(\Pi_{\tau}\vee\Pi_{\sigma})-1)}
\end{equation}
where the number 
$\gamma_{\tau\sigma^{-1},\Pi_{\tau}\vee\Pi_{\sigma}, 
q+|\tau\sigma^{-1}|+2(C(\Pi_{\tau}\vee\Pi_{\sigma})-1)}$ 
was defined in \ref{cum2} and computed in Theorem \ref{Schaeffer}.
\end{Thm}

This theorem is a consequence of the following lemma

\begin{Lem}\label{Fprop}
\begin{itemize}
\item (i)
We have, for $d\geq q$, 
\begin{equation}
E(\tr ((XUYU^*)^q))= 
\sum_{\sigma, \tau\in \mathcal{S}_q}
d^{q-|\tau |-|\sigma |}
\langle X \rangle_{\tau}\langle Y \rangle_{\sigma}\We (d,\tau\sigma^{-1})
\end{equation}
\item (ii)
\begin{equation}\label{cq}
d^{-2} C_q(d^2A)=\sum_{\sigma,\tau\in\Sy{q} } 
d^{2q-2-|\tau|-|\sigma|}
\phi_{\sigma}(X)\phi_{\tau}(Y)C_{\Pi_{\tau}\vee\Pi_{\sigma},1_q}(\tau\sigma^{-1} ,d)
\end{equation} 
\end{itemize}
\end{Lem}

\begin{demo}
The second point is a straightforward consequence of the first one together
with the definition of $C_{\Pi,1_q}(\tau\sigma^{-1},d)$.

For the first point, take the notations of \ref{eval2}:
the set $V$ is the set of odd numbers of $[1,2q]$ and the permutation $\alpha$ is
$(12)(34)\cdots (2q-1 \, 2q)$. The rest follows by inspection, together
with the fact that
$|\tau|+|\sigma|+|\tau\sigma^{-1}|$ is always an even number.
\end{demo}

Now we can prove Theorem \ref{Fborne}.

\begin{demo} 
Observe by Theorem \ref{decay}
that $C_{\Pi,1_q}(\sigma ,d)=O(d^{-q-|\sigma|-2(C(\Pi)-1)})$. 
This implies that for any $\sigma,\tau\in\Sy{q}$, one has
\begin{equation*}
d^{2q-2-|\tau|-|\sigma|}
C_{\Pi_{\sigma}\vee\Pi_{\tau},1_q}(\tau\sigma^{-1} ,d)=
O(d^{2q-|\tau|-|\sigma|-|\tau\sigma^{-1}|-2C(\Pi_{\sigma}\vee\Pi_{\tau})})
\end{equation*}
which is known to be asymptotically bounded (see \ref{decay}). 
Theorem \ref{Fborne} follows.
\end{demo}

\subsection{A geometric interpretation of the formula for the limit}

In this section we discuss briefly 
an interpretation of Theorem \ref{Fborne} in terms of sum over equivalence
class of planar graphs.
Let $G_q$ be the set of (not-necesssarily connected) planar graphs on the sphere
with $q$ edges together with the following data and conditions:

\begin{itemize}
\item (i)
each face has an even number of edges.
\item (ii)
the edges are labeled from 1 to $q$.
\item (iii)
there is a bicolouration in white and black of the vertices such that each
black vertice has only white neighbours and vice versa.
\end{itemize}

To each such graph $g\in G_q$ we associate the permutations $\sigma (g)$ 
(resp. $\tau (g)$)
of $\Sy{q}$ defined by turning clockwise (resp. counterclockwise) 
around the white (resp. black) vertices
and the function 
\begin{equation*}
\Moeb(g)=\gamma_{\tau\sigma^{-1},\Pi_{\tau}\vee\Pi_{\sigma}, 
q+|\tau\sigma^{-1}|+2(C(\Pi_{\tau}\vee\Pi_{\sigma})-1)}
\end{equation*}
For example in the picture, 
\begin{equation*}
\begin{split}
\sigma=(1 \, 13 \, 2)(3\, 5\, 4)(6 \, 7)(8\, 9 \, 10)(11 \, 12)(16 \, 17)(14 \, 15)\\
\tau=(5\, 6)(7\, 8)(10\, 11)(2\, 3\, 9)(12 \, 13)(1\, 4)(14 \, 17)(15 \, 16)\\
\tau\sigma^{-1}=(1\, 3)(5\, 9\, 7)(6\, 8\, 11\, 13\, 4)(2\, 12\, 10)(17\, 15)(14\, 16)\\
\end{split}
\end{equation*}

\begin{figure}
\centering
\epsfig{figure=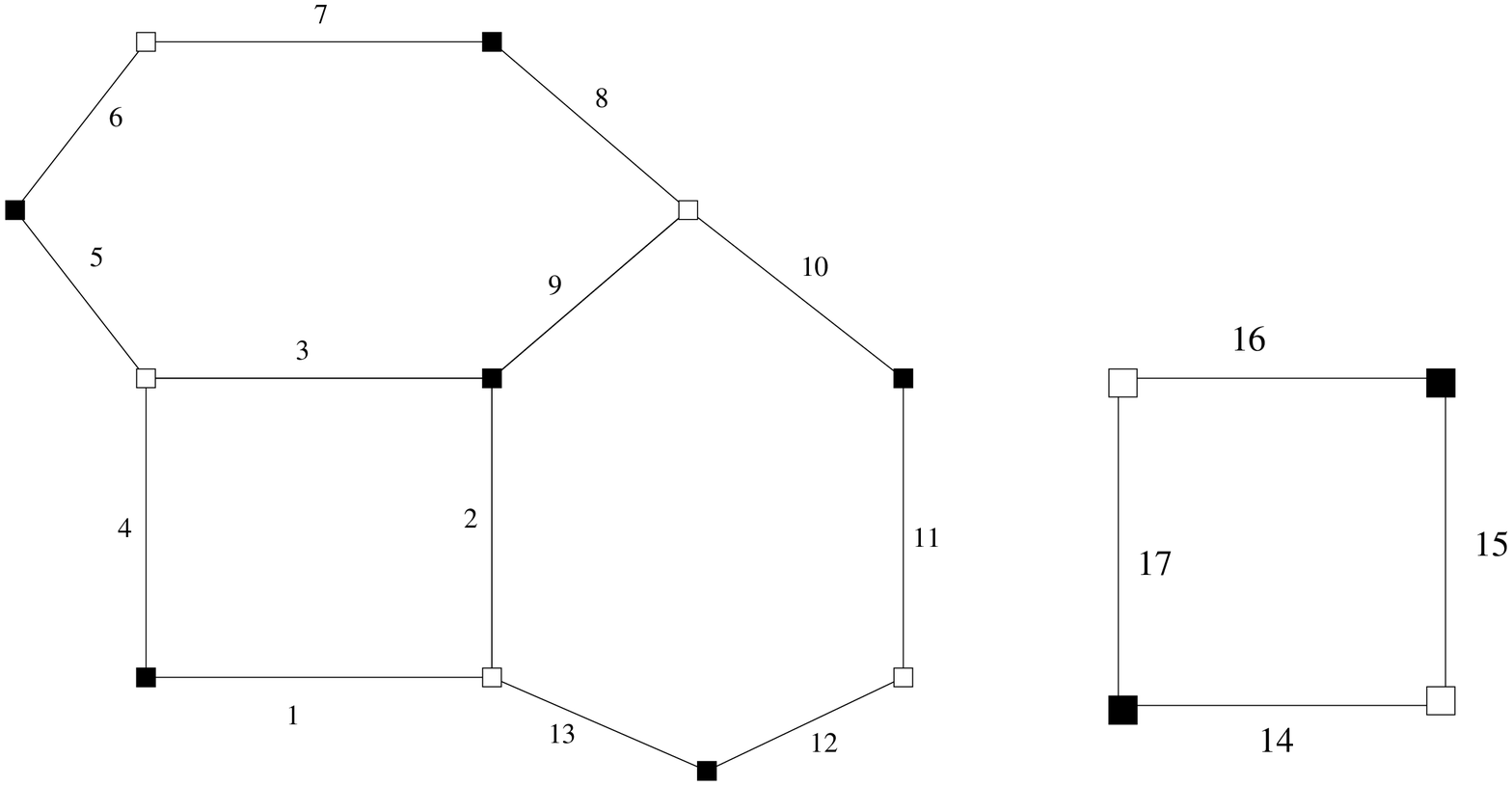, width=11cm}
\label{fig:planaire}
\end{figure}

Two graphs are said to be equivalent if there is a positive oriented diffeomorphism
of the plane transforming one to the other and respecting the colouring of
the vertices and the labeling of the edges. We call $\sim$ this equivalence
relation.

\begin{Thm}\label{diagr}
We have
\begin{equation}
\lim_d d^{-2}C_q (d^2A)=\sum_{g\in G_q/\sim}\langle X \rangle_{\tau(g)}\langle Y \rangle_{\sigma (g)}\Moeb(g)
\end{equation}
\end{Thm}

\begin{demo}
It is enough to check that our combinatorial objects of $G_q/\sim$ yield a one-to-one
encoding of the situation 
$|\sigma|+|\tau|+|\tau\sigma^{-1}|=2(q-C(\mathcal{O}_{\Pi_{\tau}}\vee\mathcal{O}_{\Pi_{\sigma}}))$.
This is a consequence of the results of \cite{MR2001g:05006}. 
\end{demo}

\subsection{A new scaling for $F_{d,X,Y}$}\label{ss4-4}

We start by recalling some definitions and facts contained in \cite{MR95h:05012}.
Let $NC_q$ be the set of non-crossing partitions of the
ordered set $[1,q]$. A partition $V=\{V_1,\ldots ,V_k \}$ of $[1,q]$ is said to
be {\it non-crossing \it} iff it is not possible to find four elements 
$a_1<a_2<a_3<a_4$ of $[1,q]$ such that $a_1$ and $a_3$ are in one block
$V_i$ and $a_2$ and $a_4$ are in one other block $V_j$, $j\neq i$.
This is a lattice for the refinement order, admitting $0_q$ as a minimal element
(the discrete partition) and $1_q$ as a maximal element (the one block 
partition). More generally, if $\Pi_1\leq\Pi_2$, $\Pi_1$ and $\Pi_2\in NC_q$,
then the interval $[\Pi_1,\Pi_2]$ is a lattice, and there exists a $\Pi\in NC_q$
such that $[\Pi_1,\Pi_2]$ is lattice-isomorphic to the interval $[O_q,\Pi]$.

Let $\sigma\in\mathcal{S}_q$ be a permutation such that $\Pi_{\sigma}=\Pi$.
The Möbius function is known (cf \cite{MR95h:05012}) to satisfy 
\begin{equation*}
\Moeb (\pi_1, \pi_2)=\Moeb (\sigma)
\end{equation*}
where $\Moeb (\sigma )$ was defined in equation \ref{defsp}.
For $A=( X_1,\cdots ,X_q )$ a $q$ -uple of
random variables 
in a non-commutative probability space $(\mathcal{A}, \phi)$, consider 
a partition $V=\{ V_1 ,\cdots ,V_k \}$,
with $V_i=\{ m_{i1}<\cdots <m_{il_i} \}$. Associate to this partition
the scalar
\begin{equation*}
A_{V}=\phi (a_{m_{11}}\cdots a_{m_{1l_1}})  \cdots 
 \phi (a_{m_{k1}}\cdots a_{m_{kl_k}})
\end{equation*}
For any integer $q$ and $\Pi\in NC(q)$, the {\it free cumulant \it} associated to $\Pi$ is the
\begin{equation}
k_{\Pi}(A)=\sum_{\Pi'\leq\Pi \, , \, \Pi'\in NC_q} A_{\Pi'}\Moeb(\Pi',\Pi)
\end{equation}
For convenience we define $k_q(A)=k_{1_q}(A)$ and
$k_q(X)=k_q(A)$ for $A=(X,\cdots , X)$.
It has been known since Speicher (see \cite{MR95h:05012}) that if
two random variables $X$ and $Y$ of a non-commutative space are free, then
for all $q\geq 1$, 
\begin{equation*}
k_q(X+Y)=k_q(X)+k_q(Y)
\end{equation*}

\begin{rem}
The elements $X\rightarrow k_q(X)$ were already known to Voiculescu under the form
\begin{equation*}
R_{X}:z\rightarrow\sum_{q\geq 0}z^q k_{q+1}(X)
\end{equation*}
as the $R$ -transform, the analytic transformation linearizing the free additive
convolution. 
\end{rem}

The following theorem is a stronger version of a result of P. Zinn-Justin
contained in \cite{MR2000h:82043} (also see \cite{MR2000e:82029a}). 

\begin{Thm}\label{thmfree}
Let $X_d$ be a rank one projector and assume that $Y_d$ has a limit distribution.
\begin{equation}\label{final}
\lim_d d^{-1}\cdot C_q(d^2A_d)=(q-1)!k_q(Y) 
\end{equation}
\end{Thm}

In order to prove this theorem we need some notation and some preliminary results.
For a partition $V=\{V_1,\cdots ,V_k\}$ of $[1,q]$, enumerate 
the elements of $V_i$ as an increasing sequence $v_{i1}\leq\cdots\leq v_{il_i}$.
Then we define the right converse $r$ of the surjection by its cycle
product decomposition:
\begin{equation*}
r(V)=\sigma = (v_{11}\cdots v_{1l_1})\cdots (v_{k1}\cdots v_{kl_k})
\end{equation*}
This injection satisfies $\Pi_{r(V)}=V$.

\begin{Lem} \label{NC}
Let $Z$ be the permutation $(1 \cdots q)$ of $\mathcal{S}_q$.
The set of $\tau$ 's satisfying
\begin{equation*} 
|\tau|+|Z\tau^{-1}|=q-1
\end{equation*}
is in one to one bijection with the set $NC(q)$. 
This bijection is $\tau\rightarrow\Pi_{\tau}$  
and its converse is $r$.
\end{Lem}

\begin{demo}
See paragraph 2.7 of \cite{MR2001b:05225}, or \cite{MR98h:05020}.
\end{demo}

The {\it Kreweras complementation \it} $K(\pi)$
is defined, for $\pi$ a non-crossing partition of $NC_q$, as
the partition $\Pi_{Z^{-1} r(\Pi)}$, with $Z=(1\cdots q)$.
We first establish a formula for the free cumulant, slightly different
from the usual ones.

\begin{Lem}\label{speicher}
One has
\begin{equation} 
k_q(Y)=\sum_{p\in NC(q)} \phi_{p}(Y) \cdot \Moeb (K(p))
\end{equation}
\end{Lem}

\begin{demo}
The interval $[p,1_q]$ is isomorphic as a lattice to the interval
$[0,K(p)]$ because the Kreweras complementation is an antiautomorphism
of the lattice of non-crossing partitions.
Consequently we get that $\Moeb (p,1_q)=\Moeb (0_q,K(p))=
\Moeb (K(p))$ and thus:
\begin{equation*}
k_q(Y)=\sum_{p\in NC_q} \phi_p (Y) \cdot \Moeb (p,1_q)=
\sum_{p\in NC_q} \phi_p (Y) \cdot\Moeb (K(p))
\end{equation*}
\end{demo}

Now we can proceed to the proof of theorem \ref{thmfree}.

\begin{demo}
Recall from \ref{cq} that 
\begin{equation*}
d^{-1} C_q(d^2A)=\sum_{\sigma,\tau\in\Sy{q} } 
d^{1+2q-|\tau|-|\sigma|-|\tau\sigma^{-1}|-2}
\phi_{\sigma}(X)\phi_{\tau}(Y)C_{\Pi_{\sigma}\vee\Pi_{\tau} ,1_q}(\tau\sigma^{-1} ,d)
\end{equation*}
In particular here 
$\phi_{\sigma}(X)=d^{-q+|\sigma|}$, so every summand of Equation \ref{cq}
is $O(d^{x})$ with $x=2q-|\tau|-|\sigma|-|\tau\sigma^{-1}|-2C(\Pi_{\sigma}\vee\Pi_{\tau})
-q+|\sigma|+1$.
We already know that for all permutations $\sigma$ and $\tau$ we have
$2q-|\tau|-|\sigma|-|\tau\sigma^{-1}|-2C(\Pi_{\sigma}\vee\Pi_{\tau})\leq 0$;
and it is also clear that  $-q+|\sigma|+1\leq 0$.
If the exponent $x$ is zero then $|\sigma|$ has to be $q-1$.
In such a case $\Pi_{\sigma}\vee\Pi_{\tau}=1_q$ and
$x=q-1-|\tau|-|\tau\sigma^{-1}|$. The permutation $\sigma$ has to be
conjugated to the permutation $Z=(1\ldots q)$ and its conjugacy class
has $(q-1)!$ elements. Therefore
\begin{equation}\label{cqq}
d^{-1} C_q(d^2A)=(q-1)!\sum_{\tau\in\Sy{q},|\tau|+|\tau Z^{-1}|=q-1 } 
\phi_{\tau}(Y)\We(\tau\sigma^{-1} ,d)+o(1)
\end{equation}
Lemmas \ref{speicher} and \ref{NC} imply that 
Equation \ref{cqq} is the same as Equation \ref{final}, which is the
expected result.
\end{demo} 

\section{Conclusion and numerical values}\label{s-5}

\subsection{Conclusion}

The results of this paper
raise a number of questions: Is there any chance to obtain 
a reasonably short explicit formula for the limit in Theorem
\ref{Fborne}, for example in terms of
some generating function? In this direction, P. Zinn-Justin recently communicated me
a paper (\cite{pzj}) in which he computes the coefficients
of $d^{-2}C_q(d^2\tr A)$ of the kind $P(y)x_{i_1}x_{i_2}$ and $P(y)x_{i_1}x_{i_2}x_{i_3}$
(see \ref{ss-5.3} for notation).

In another direction, does our result hold in the topology of uniform
convergence on compact subsets of $\mathbb{C}$ ? This question was the
one we hoped to answer in view of the results of \cite{MR1883414}, but it still
remains open to us. 

Finally, we believe that our approach to integration of polynomial functions on 
unitary groups could lead to a better understanding 
of statistical properties of eigenvalues for the product of independent 
unitary random variables. For example, if $U$ and $V$ are independent 
unitary random variables of $\M{d}$, we obtain thanks to lemma \ref{eval2}:
\begin{equation*}
\begin{array}{lll}
E(\tr(UVU^*V^*))=d^{-2} &
E(\tr(U^2VU^{*2}V^*))=2d^{-2} \\
E(\tr(UVUVU^*V^*U^*V^*))=2d^{-2} &
E(\tr(U^2V^2U^{*2}V^{*2}))=\frac{3d^2-4}{d^2(d^2-1)} \\
E(\tr(U^3VU^{*3}V^*))=3d^{-2} & E(\tr((UVU^*V^*)^2)=\frac{-4}{d^2(d^2-1)}\\
\end{array}
\end{equation*}
An elementary use of symmetric properties of the algebra generated by
$U$ and $V$ shows that the above list allows one to compute the expectation of
the trace of all reduced words in $U,V,U^*,V^*$ of length
eight or less. The method can be generalized to arbitrary length words.
 
\subsection{Values of $\We$ for small $q$.}\label{ss-5.1}

We give a table of the first $\We$ functions.
Note that the asymptotics of these rational fractions
fit with Theorem \ref{Schaeffer}, (iii):

\begin{equation*}
\begin{array}{lll}
\We((1),d)=\frac{1}{d} &
\We((2),d)={\frac {-1}{d\left (d^2-1\right )}}\\
\We((1,1),d)={\frac {1}{\left (d^2-1\right )}} &
\We((3),d)={\frac {2}{\left (d^2-1\right )\left (d^2-4\right )
d}}\\
\We((2,1),d)={\frac {-1}{\left (d^2-1\right )\left (d^2-4\right )
}} &
\We((1,1,1),d)={\frac {{d}^{2}-2}{\left (d^2-1\right )\left (d^2-4
\right )d}}\\
\end{array}
\end{equation*}

We also observe that 
\begin{equation}
\We((q),d)=(-1)^{q+1}c_{q-1}\prod_{-q+1\leq j \leq q-1}(d-j)^{-1}
\end{equation}
Indeed, a classical combinatorial result about the Schur polynomials
shows that if one writes $\We((q),d)$ as an irreducible rational fraction,
its denominator has to be $\prod_{-q+1\leq j \leq q-1}(d-j)$. An argument
of degree of $\We((q),d)$
together with the knowledge of the asymptotics concludes the proof.

\subsection{Values of $\lim_d d^{-2}C_q(d^2A)$ for small $q$.}\label{ss-5.3}

We give an asymptotic value of the first cumulants. The four first values are due to
Biane and we performed others with a computer.
Since they are semi-invariant, it is enough to assume that $\tr X= 0$ and $\tr Y=0$.
This avoids (even more) cumbersome results.
We use the following notation:
$A=XUYU^*$. 
$\lim_d\tr X^i = x_i$ and $\lim_d\tr Y^i=y_i$.

\begin{equation*}
\begin{split}
d^{-2}C_1(d^2\tr A)/0!=0\\
d^{-2}C_2(d^2\tr A)/1!=x_{{2}}y_{{2}}\\
d^{-2}C_3(d^2\tr A)/2!=x_{{3}}y_{{3}}\\
d^{-2}C_4(d^2\tr A)/3!=x_{{4}}y_{{4}}+3\,{x_{{2}}}^{2}{y_{{2}}}^{2}-2\,y_{{4}}{x
_{{2}}}^{2}-2\,x_{{4}}{y_{{2}}}^{2}\\
d^{-2}C_5(d^2\tr A)/4!=x_{{5}}y_{{5}}+20\,x_{{2}}x_{{3}}y_{{2}}y_{{3}}-5\,x_{{
2}}x_{{3}}y_{{5}}-5\,x_{{5}}y_{{2}}y_{{3}}\\
d^{-2}C_6(d^2\tr A)/5!=-6\,y_{{6}}x_{{2}}x_{{4}}+30\,x_{{2}}x_{{4}}y_{{2}}y_{{4}}
+27\,{x_{{2}}}^{3}{y_{{2}}}^{3}-30\,{x_{{2}}}^{3}y_{{2}}y_{{4}}\\
-
30\,{y_{{2}}}^{3}x_{{2}}x_{{4}}-16\,{y_{{2}}}^{3}{x_{{3}}}^{2}+
15\,x_{{2}}x_{{4}}{y_{{3}}}^{2}+15\,{x_{{3}}}^{2}y_{{2}}y_{{4}}\\
-
6\,x_{{6}}y_{{2}}y_{{4}}+x_{{6}}y_{{6}}-16\,{x_{{2}}}^{3}{y_{
{3}}}^{2}+7\,x_{{6}}{y_{{2}}}^{3}+7\,{x_{{2}}}^{3}y_{{6}}\\
+6\,{x_
{{3}}}^{2}{y_{{3}}}^{2}-3\,y_{{6}}{x_{{3}}}^{2}-3\,x_{{6}}{y_{{3}}
}^{2}\\
d^{-2}C_7(d^2\tr A)/6!=-7\,y_{{7}}x_{{2}}x_{{5}}+42\,x_{{2}}x_{{5}}y_{{3}}y_{{4
}}+42\,x_{{3}}x_{{4}}y_{{2}}y_{{5}}-7\,x_{{7}}y_{{2}}y_{{5}}\\
+
462\,x_{{3}}y_{{3}}{x_{{2}}}^{2}{y_{{2}}}^{2}-7\,x_{{7}}y_{{3}}y
_{{4}}+x_{{7}}y_{{7}}+28\,x_{{7}}{y_{{2}}}^{2}y_{{3}}\\
-7\,x_
{{3}}x_{{4}}y_{{7}}+35\,x_{{3}}x_{{4}}y_{{3}}y_{{4}}-140\,x_{{3}
}x_{{4}}{y_{{2}}}^{2}y_{{3}}+28\,{x_{{2}}}^{2}x_{{3}}y_{{7}}\\
-140
\,{x_{{2}}}^{2}x_{{3}}y_{{3}}y_{{4}}+42\,x_{{2}}x_{{5}}y_{{2}}y_{{5
}}-147\,x_{{2}}x_{{5}}{y_{{2}}}^2y_3-147\,x_2^2x_3y_2y_5\\
\end{split}
\end{equation*}

These computations have been performed with Maple and its ACE package
on a Free BSD Dell personal computer at the École Normale Supérieure.

\bibliographystyle{alpha}
\bibliography{article1201}

\end{document}

%% file: notations.tex
\newcommand{\M}[1]{\mathbb M_{#1}(\mathbb{C})}
\newcommand{\Sy}[1]{\mathcal{S}_{#1}}

\newcommand{\Moeb}{{\rm Moeb \rm}}

\newcommand{\Tr}{{\rm Tr \rm}}
\newcommand{\tr}{{\rm tr \rm}}
\newcommand{\card}{{\rm Card \rm}}
\newcommand{\We}{{\rm Wg \rm}}

\newtheorem{Thm}{Theorem}[section]
\newtheorem{Prop}[Thm]{Proposition}
\newtheorem{Lem}[Thm]{Lemma}
\newtheorem{Cor}[Thm]{Corollary}

\newtheorem*{Thm*}{Theorem}
\newtheorem*{Cor*}{Corollary}

\newcounter{ploum}
\newcounter{ex}[section]
\newcounter{rem}[section]
\newcounter{ass}[section]

\numberwithin{equation}{section}